\documentclass[utf8,final]{frontiersSCNS}
\usepackage{url,lineno,microtype,subcaption}
\usepackage[breaklinks]{hyperref}
\usepackage[utf8]{inputenc}
\usepackage{graphicx}
\usepackage{color}
\usepackage{amsmath,amssymb,amsfonts}
\usepackage{bm,bbm,multirow}
\newcommand{\bi}{\begin{itemize}}
\newcommand{\ei}{\end{itemize}}
\newcommand{\ben}{\begin{enumerate}}
\newcommand{\een}{\end{enumerate}}
\newcommand{\be}{\begin{equation}}
\newcommand{\ee}{\end{equation}}
\newcommand{\ba}{\begin{eqnarray}}
\newcommand{\ea}{\end{eqnarray}}
\def\bs{\begin{subequations}}
\def\es{\end{subequations}}
\def\a{\alpha}
\def\b{\beta}
\def\de{\delta}

\def\la{\lambda}

\def\om{\omega}
\def\De{\Delta}

\def\s{\sigma}

\newcommand{\Eq}[1]{(\ref{#1})}

\def\cob{\color{blue}}

\renewcommand{\leq}{\leqslant}
\renewcommand{\geq}{\geqslant}

\newcommand{\oarX}[1]{\href{http://arxiv.org/abs/#1}{{\ttfamily\cob arXiv:#1}}}
\newcommand{\arX}[1]{\href{http://arxiv.org/abs/#1}{{\ttfamily\cob arXiv:#1}}}

\newcommand{\doin}[6]{\href{http://dx.doi.org/#1}{{\cob {\it #2 #3} {\it #4}, #5}}}
\newcommand{\doinn}[5]{\href{http://dx.doi.org/#1}{{\cob {\it #2} {\it #3}, #4}}}
\newcommand{\doinny}[5]{\href{http://dx.doi.org/#1}{{\cob {\it #2} {\it #3}, #4 (#5)}}}

\newcommand{\ndoinn}[5]{\href{#1}{{\cob {\it #2} {\it #3}, #4}}}
\newcommand{\doiee}[4]{\href{http://dx.doi.org/#1}{{\cob {\it #2, #3}}}}

\newcommand{\procsing}[7]{#2 (Ed.), \emph{#1} (pp.~#7). #4, #5: #3}
\newcommand{\procmany}[7]{#2 (Eds.), \emph{#1} (pp.~#7). #4, #5: #3}
\newcommand{\procmanys}[6]{#2 (Eds.), \emph{#1}. #4, #5: #3}
\newcommand{\procmanyd}[4]{#3 (Eds.), \href{#1}{\cob\it #2} (pp.~#4)}
\newcommand{\book}[5]{\emph{#1}. #3, #4: #2}
\newcommand{\tia}[1]{#1.}
\def\rme{e}
\def\rmd{d}

\def\keyFont{\fontsize{8}{11}\helveticabold }
\def\firstAuthorLast{Calcagni {et~al.}} %use et al only if is more than 1 author
\def\Authors{Gianluca Calcagni\,$^{1,*}$, Ernesto Caballero-Garrido\,$^{2}$ and Ricardo Pellón\,$^{3}$}
\def\Address{$^{1}$Instituto de Estructura de la Materia, CSIC, Madrid, Spain\\
$^{2}$National Association of Researchers, Twenty-First Century, Madrid, Spain\\
$^{3}$Facultad de Psicología, UNED, Madrid, Spain}
\def\corrAuthor{Gianluca Calcagni}

\def\corrEmail{g.calcagni@csic.es}

\begin{document}
\onecolumn
\firstpage{1}

\title[Behavior stability in Pavlovian conditioning]{Behavior stability and individual differences in Pavlovian extended conditioning} 

\author[\firstAuthorLast]{\Authors} %This field will be automatically populated
\address{} %This field will be automatically populated
\correspondance{} %This field will be automatically populated

\extraAuth{}

\maketitle

\begin{abstract}
\noindent How stable and general is behavior once maximum learning is reached? To answer this question and understand post-acquisition behavior and its related individual differences, we propose a psychological principle that naturally extends associative models of Pavlovian conditioning to a \emph{dynamical oscillatory model} where subjects have a greater memory capacity than usually postulated, but with greater forecast uncertainty. This results in a greater resistance to learning in the first few sessions followed by an over-optimal response peak and a sequence of progressively damped response oscillations. We detected the first peak and trough of the new learning curve in our data, but their dispersion was too large to also check the presence of oscillations with smaller amplitude. We ran an unusually long experiment with 32 rats over 3960 trials, where we excluded habituation and other well-known phenomena as sources of variability in the subjects' performance. Using the data of this and another Pavlovian experiment by \cite{HPG}, as an illustration of the principle we tested the theory against the basic associative single-cue Rescorla--Wagner (RW) model. We found evidence that the RW model is the best nonlinear regression to data only for a minority of the subjects, while its dynamical extension can explain the almost totality of data with strong to very strong evidence. Finally, an analysis of short-scale fluctuations of individual responses showed that they are described by random white noise, in contrast with the colored-noise findings in human performance.

\tiny
 \keyFont{\section{Keywords:} Pavlovian conditioning, Rescorla--Wagner model, Associative models, Extended training, Individual differences, Bayes information criterion, $1/f$ noise}
\end{abstract}

\date{Original: May 27, 2018; revised: March 4, 2020}

%\tableofcontents

%%%%%%%%%%%%%%%%%%%%%%%%%%%%%%%%%%%%%%%%%%%%%%%%%%%%%%%%%%%%%%%%%%%%%%%%%%%%%%
%%%%%%%%%%%%%%%%%%%%%%%%%%%%%%%%%%%%%%%%%%%%%%%%%%%%%%%%%%%%%%%%%%%%%%%%%%%%%%

\section{Introduction}%\section{INTRODUCTION}

How stable is behavior when there is nothing more to learn? Much debate has been flourished around this basic question since the earliest studies of animal conditioning \citep{Pav27}, especially after the first efforts to make the discipline theoretically quantitative with a mathematical approach \citep{Hul43}. Observations point towards an instability of the response in extended training. In the context of discrimination experiments of operant conditioning, extended training was studied in relation with behavioral contrast and the peak-shift effect. Pigeons trained for 60 days with interspersed generalization testing showed a gradual response decrease \citep{Te66a}. In an experiment lasting 64 sessions, \cite{Hea71} did not observe this decrease from peak responding, sometimes called {overtraining effect} (as a reduction in behavioral contrast), {inhibition with reinforcement}, or {post-peak depression} (see \citealp{KiBu}, for an early review and other references). Extending the training to 105-125 days, the response decrease was found to be a subject-dependent and transient effect, giving way to a greater variety of patterns characterized by apparently random response fluctuations and, in general, remarkable individual differences \citep{DuLy}. An attenuated conditioned responding with extended reinforced training has been observed also in the case of Pavlovian conditioning, where it is modulated by the context \citep{Bou08,OPBLL,UWM}. Post-peak depression is a rather short-scale phenomenon usually achieved within a few sessions and not too many trials. For instance, the experiments with dogs by \cite{OPBLL} showed response decrease on a time scale of 300 trials. In the experiments of \cite{Bou08}, the groups of subjects received about two weeks of training. In the case of \cite{UWM}, they were given 5 to 6 sessions of 5 to 60 trials, for a maximum of about 360 trials. For acquisition of fear conditioning, 100 pairings during 10 days are sufficient \citep{Pic09}. However, longer-term cases are known, such as the first documented case of inhibition with reinforcement. Pavlov (\citeyear{Pav27}, Lecture XIV) reported experiments with a dog that spanned several years and that showed a progressive decrease in the conditioned response when the same type of stimuli was applied. On the other extreme of the spectrum, response fluctuations have been registered also on the very short time scale of trial by trial \citep{ABKMS}.

The prototypical learning curve of Pavlovian conditioning in the presence of a single cue was described by \citet{Hul43}. \citet{Est50} and \citet{BuMo1} wrote down a mathematical model (somewhat implicit in Hull's discussion) in terms of response probability, but for operational reasons the latter was replaced by the association strength $v$ by \citet{RW72}. Due to this complicated genesis, the resulting single-cue model in terms of $v$ has received several names: Hull, Hull--Spence, Estes, Bush--Mosteller, and single-cue Rescorla--Wagner, among others \citep{LeP,WV}. For brevity, we will call it \emph{Rescorla--Wagner (RW) model} here.

Including RW, most conditioning models are about learning, which means that their simulation of the execution or reaction of the subject once the asymptote is reached has not been validated extensively. A classic problem consists in that, when one has learned everything, it is not convenient to keep giving attention to the stimuli of the task and there is a transition to a more automatic mode of execution. In order to explain this transition, many Pavlovian models (e.g., the Pearce--Hall model, \citeyear{PH}) distinguish between automatic and controlled processing. Still, this difference plays a role in the first few sessions of training and it does not address the issue of what happens after thousand of trials. Going beyond associative models, the opponent-processes theory \citep{SoCo} and the SOP model \citep{Wag81} provide a partial, but not entirely comprehensive, explanation of post-peak depression and related phenomena.

These studies also highlight the parallel issue of individual differences. Individual plots are so non-smooth and erratic that any vestige of the clean, smooth learning curve of averaged data may be completely lost. When averaging, information on individuals is usually lost. This concern is not new and it was voiced already in early days of the discipline \citep{Hay53,Mer31,Sid52} and retaken into consideration in recent years (\citealp{BlMo,Gal12,Gal04,Gla13,JVFF,You18}; see especially \citealp{SL18}). As \cite{Sid52} pessimistically put it, ``[i]ntra-organism variability may be so great as to obscure any lawful relation.'' Smooth group-learning curves have even been stigmatized as an artifact, since step-like sudden acquisition has been observed in several experiments \citep{Gal04}. Despite these warnings, however, averaging the data can be a useful procedure \citep{Est56} and is still commonly employed in the great majority of publications, even those where individual responses are analyzed \citep{MaHa}.

All this literature helps to refocus the question we proposed in the opening and to give the term ``stability'' two different meanings. One corresponds to intra-subject behavior stability: the variability of the individual response throughout the experiment. The other is inter-subject stability, in the sense of the range and variety of patterns that individual differences can take when the performance of experimental subjects is compared. Both intra- and inter-subject stability can refer to phenomena spanning trials (short-term stability) or sessions (long-term stability). Short-term stability usually pertains to the initial acquisition stage of conditioning, where the subjects are in the process of acquiring maximal learning but have not quite reached the asymptote of their learning curve. Long-term stability is more related to response at the asymptote. Response variations in the form of random fluctuations may be regarded both as short-term effects (they occur as gradients from one session to another) and as long-term, since they can span several sessions (or when one detects oscillation-like features with a long period).

To the best of our knowledge, the mainstream of mathematical associative models starting from RW predicts an indefinitely long asymptotic permanence of execution in the learning process, for each and any subject. Neither individual differences nor response fluctuations are considered in most analytic treatments of the theories, notwithstanding the number of exceptions to this generalized trend, some of which we have mentioned above. An inversion of this trend has been seen recently, when new models have arisen that give more importance to individual differences. The multiple-state learning model of \cite{BlMo} and the MECA model of \cite{Gla13} are examples. Also, Estes' stimulus sampling theory \citeyearpar{Est50} is one of the earliest attempts to quantify and explain variability in learning progress, as due to fluctuations in environmental and internal factors. The issue at stake here is not just whether there exist superior \emph{ad hoc} fits to averaged data than that provided by the RW model with one cue. It is already known that other types of learning curves can fare better than the exponential profile (see Eq.\ \Eq{Hullearn} below), even at the individual level. A power-law curve \citep{NeRo} or the accumulation model \citep{MaHa} are two instances. Rather, here we are interested in the problem of stability in the double sense specified above and, moreover, any new model should arise as an underlying theory rather than just a tailor-made learning curve.

With the aim to study both very long-term stability and individual differences, we present the results of an experiment of Pavlovian conditioning that ran through a total of almost 4000 trials. The first goal of this paper is to check how variable is behavior in the long-term post-acquisition phase. We do find fluctuations around the asymptote, both on a trial-by-trial and a session-by-session basis, but not statistically significant. In other words, behavior is fairly stable even when the subject is no longer learning. This provides a validation of the RW single-cue associative model even in the not-so-often explored plateau region of the learning curve, far away from initial acquisition. However, the fit of the data of individual subjects is much more unstable and one may wonder whether there exist a quantitative model accounting for this variability. Our second goal is to explore several associative models extending RW's in a most natural way, introducing a psychological principle that, in analogy with the same tool used in classical\footnote{To avoid an otherwise inevitable confusion, in this paper we use the term ``classical'' always in the sense of physics, calling the traditional associative models of conditioning ``Pavlovian.''} mechanics physics, we will call of \emph{least action}. This principle states that learning processes must be described as dynamical systems, where \emph{dynamical} means that there exists a quantity (the \emph{action}) that must be minimized when the association strength is changed during conditioning. %Independently, we will also introduce a model governed by a random noise.
 Through a detailed statistical and spectral analysis, we show that, despite the large fluctuations of the subjects' behavior, the RW model is still a good fit to data, except in 4 out of 15 experimental subjects. The \emph{dynamical oscillatory model} (or DOM in short), based on the principle of least action and predicting resistance to learning in the first sessions, provides a better fit to the data of this 20\% of the sample, according to both Bayesian and Akaike Information Criteria. Since the DOM is an extension of RW rather than a competitor, we conclude that it fits successfully 100\% of the data, while the RW model accounts only for 80\%. These results are strengthened when analyzing the data of Experiment 2 of \citet{HPG}, where the RW model can fit successfully only 45\% of the data and the DOM 95\%. Although the existence of a learning asymptote has been questioned in the past \citep{Gal04}, we find its presence to be a robust feature of all the models. The formalism we propose can be extended to the multi-cue and varying-salience cases and can account for the non-normative performance of some subjects.

The RW model may be regarded as antiquated today. Context is not fully integrated into it and few learning preparations can be viewed as single-cue because learning does not occur in a vacuum \citep{MiMa}. Furthermore, the inability of the Rescorla--Wagner model to account for recovery from extinction and other phenomena is well known \citep{MBG}. Nevertheless, we believe that the novelty of the least-action principle and the ensuing dynamical proposal can be better understood and appreciated if compared with the simplest representative of the standard canon. As we will see when extending the theory to many cues, there seems to be no obstacle to apply the principle to more realistic situations.

One of the main features of this work is that traditional associative models are not extended in an \emph{ad hoc} way, but by using a rigorous top-down procedure leading to a natural (in the sense of logic-based) conclusion, closer to a theory rather than a phenomenological model. To that aim, we need mathematics more advanced than those available to a large portion of the readership in psychology, but the payback offered in terms of explanation of the data may be worthwhile. In order to keep the presentation simple, we will introduce the model in a pedagogical way, confining the most rigorous parts and other dynamical models to the \textbf{Supplementary Material}.

The oscillations of the DOM span tens of sessions and describe individual variations in the response at large time scales. In order to study short-scale (session-by-session or even trial-by-trial) variations, we need other models and analysis tools. One such model treats short-scale fluctuations as random noise and is capable to determine whether their origin is just experimental uncertainty or some deeper, perhaps cognitive, mechanism.

The plan of the article is the following. We first review the RW associative model in section \ref{Hull model} \emph{Single-cue RW model} and clarify whether it should be applied to individual subjects or to their average. In order to clarify the type of phenomena we would like to explore, in section \ref{two1} \emph{A long experiment on Pavlovian conditioning}, we present a 3960-trial-long experiment, with a first analysis centered on the average learning curve, while in section \ref{two2} \emph{Experiment 2 of Harris et al.\ (2015)} we reanalyze the original data of one of the preparations of \citet{HPG}. Two alternative models of individual conditioning are discussed in sections \ref{dyns} \emph{Dynamical models of individual behavior} (a general framework where we reformulate the RW model and introduce a new model where subjects initially show resistance to learning) and \ref{noises} \emph{Colored stochastic model of individual behavior}, where we formulate a descriptive model of short-scale random fluctuations of individual responses and contrast it with the data. Final remarks are collected in section \ref{geco} \emph{Conclusions}. The \textbf{Supplementary Material} is devoted to material that would disrupt the flow of the main text.

%%%%%%%%%%%%%%%%%%%%%%%%%%%%%%%%%%%%%%%%%%%%%%%%%%%%%%%%%%%%%%%%%%%%%%%%%%%%%%
%%%%%%%%%%%%%%%%%%%%%%%%%%%%%%%%%%%%%%%%%%%%%%%%%%%%%%%%%%%%%%%%%%%%%%%%%%%%%%

\section{Single-cue RW model}\label{Hull model}%\section{HULL MODEL: SINGLE-CUE RESCORLA--WAGNER}

\begin{figure}[!ht]
\includegraphics[width=8.5cm]{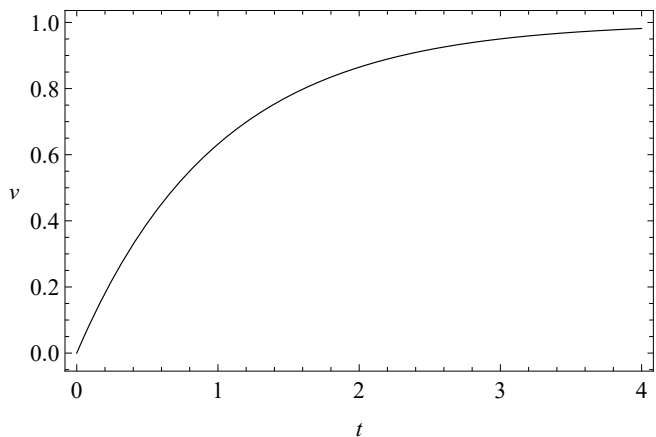}
\caption{\label{fig1} Learning curve: the solution \Eq{sol1} $v(t)$ ({\it vertical axis}) of RW conditioning model \Eq{DeV2} as a function of time $t$ ({\it horizontal axis}), for $c=\la=1$ and $\a\b=1$ (excitatory conditioning).}
\end{figure}

According to the model developed by Rescorla and Wagner, the association between the conditioned stimulus (CS) and the unconditioned stimulus (US) in Pavlovian training can be measured, at the $n$-th trial or session, by the operational variable $v_n$, called association strength. Usually in the literature, this is denoted with $V_n$, but here we use a small letter to avoid confusion with the potentials introduced below. The change $\De v_n:=v_n-v_{n-1}$ in the strength of the association at the $n$-th trial is
\be\label{DeV}
\De v_n=\a\b(\la-v_{n-1})\,,\qquad n=1,2,3,\dots,
\ee
where $0\leq\a\leq 1$ is the salience of the CS, $0\leq\b\leq 1$ is the salience of the US, and $0\leq\la\leq 1$ is the magnitude of the US. For convenience, we promote the trial sequence $n=1,2,3,\dots$ to a continuous time process described by a continuous time variable $t$. This approximation is valid as long as we consider many trials or sessions. In this way, we can recast Eq.\ \Eq{DeV} as the first-order differential equation
\be\label{DeV2}
\dot v=\a\b(\la-v)\,,
\ee
where a dot denotes a derivative with respect to time, $\dot v:=\rmd v(t)/\rmd t$. Its general solution is
\be\label{sol1}
v(t)=\la-c\rme^{-\a\b t}\,,
\ee
where $c$ is a constant. The initial condition at $t=0$ of this solution is $v(0)=\la-c$, while $v(+\infty)=\la$. Therefore, for excitatory conditioning $c=\la>0$ ({\bf Figure \ref{fig1}}), while for extinction $\la=0$ and $c<0$, so that $v=|c|\exp(-\a\b t)$ decreases in time.
%1

In this article, we will consider only excitatory conditioning and attempt, among other possibilities, to fit data with the monotonic learning curve
\be\label{Hullearn}
v_\textrm{excit}(t)=\la\left(1-\rme^{-\a\b t}\right)\,.
\ee
This theoretical curve has two free parameters: $\lambda$ and the product $\a\b$. In the experiment below, we will not be able to determine the salience of the CS and US separately.

At the risk of stating the obvious, it should be noted that there are two ways in which to interpret Eqs.\ \Eq{DeV2} and \Eq{Hullearn}. One is as an associative model for individuals, in which case $v(t)$ is the association strength at a given time $t$ of a given subject. If the RW model were a reliable description of reality, all subjects should obey the model with reasonable accuracy and differ in their behavior only in the value of the parameters $\la$, $\a$, and $\b$ in the aforementioned equations. However, this interpretation is too restrictive and does not allow for individual differences in the learning process, something that any experimentalist would recognize as inevitable. However, if the majority of subjects obeyed the RW model, then the latter could be regarded as valid in average, in which case we will make it explicit that the association strength appearing in Eqs.\ \Eq{DeV2} and \Eq{Hullearn} should be replaced by the average $\langle v\rangle:=\sum_{i=1}^N v_i/N$ over the subjects:
\be\label{DeV2av}
\dot{\langle v\rangle}=\a\b(\la-\langle v\rangle)\,,\qquad \langle v\rangle_\textrm{excit}(t)=\la\left(1-\rme^{-\a\b t}\right)\,.
\ee

As we will see in this article, the RW model is a good description of Pavlovian learning both for individuals and in average. However, we do find individual differences which can be better described by an extension of the model which we will dub ``dynamical,'' and which does not improve averaged data significantly. Therefore, we will strictly keep the distinction between models for individuals ($v$) and in average ($\langle v\rangle$).

%%%%%%%%%%%%%%%%%%%%%%%%%%%%%%%%%%%%%%%%%%%%%%%%%%%%%%%%%%%%%%%%%%%%%%%%%%%%%%
%%%%%%%%%%%%%%%%%%%%%%%%%%%%%%%%%%%%%%%%%%%%%%%%%%%%%%%%%%%%%%%%%%%%%%%%%%%%%%

\section{Two experiments on Pavlovian conditioning}\label{two}%\section{TWO EXPERIMENTS ON PAVLOVIAN CONDITIONING}

\medskip

\subsection{A long experiment on Pavlovian conditioning}\label{two1}

\medskip

\subsubsection{Subjects and materials}

We employed 32 male Wistar Han rats, without food or water restriction, divided into two experimental groups (Group 1 and Group 2) and two control groups. The US was drops of saccharin solution at 0.1\% concentration for Group 1 and at 0.2\% for Group 2. One experimental subject was removed due to poor health. The US was delivered in individual conditioning boxes via a water pump activated by an electrovalve. %In the case of experimental subjects, delivery happened on a variable-time 5 s schedule (VT-5) implemented as a uniform random distribution during the emission of a 10 s tone (CS). Licks were automatically recorded. We took 90 sessions each made of 44 trials.
 For full details of the subjects and the materials, including a justification of the chosen saccharin concentrations, see the \textbf{Supplementary Material}.

All care and experimental procedures were in accordance with the Spanish Royal Decree 53/2013 regarding the protection of experimental animals and with the European Union Council Directive 2010/63. UNED bioethics committee approved the experimental protocol.

\medskip

\subsubsection{Experimental design}

The experiment was divided into three phases. In the first phase of \emph{pre-training}, subjects were exposed to the basic functioning of the liquid dispenser in the conditioning boxes. Drops of saccharin solution of the concentration corresponding to the rat's group were delivered according to a variable-time 5 s schedule (VT-5), implemented as a uniform random distribution between 3 and 7 s with steps of 1 s. Each lick was reinforced by the delivery of another drop via a fixed-ratio schedule (FR-1). Each session of pre-training lasted 10 minutes and was preceded by 30 s of darkness, lasting in total 10'30''.

\begin{figure}[!ht]
\includegraphics[width=8.5cm]{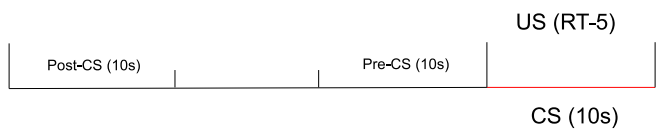}
\caption{\label{fig2} Structure of a trial for the experimental groups of our long experiment as described in the text.}
\end{figure}

The second phase (\emph{training}) consisted in sessions of total duration of 2259 s (about 37'40''). After 30 s of darkness, each box was lighted and the session went through for 44 trials, ending with 10 s of inactivity. {\bf Figure \ref{fig2}} is a scheme of a trial for the experimental groups. An inter-trial interval (ITI) of variable length averaging 40 s, realized by a uniform random distribution between 20 and 60 s with steps of 4 s, was followed by the CS, a tone of 85 db, 600 Hz, and a fixed 10 s duration. The intensity of the tone was well above the average ambient noise inside each box (65 db).%\footnote{We thank Ralph Miller for tips on the rats-and-sounds topic: ``I have found that anything between 300 and 2000 Hz works similarly if presented at least 8 dB above background'' (private communication, 2017).}
 During the CS, a US consisting in one drop of saccharin solution was delivered at random intervals of 5 s (RI-5). A random-interval schedule \citep{Mil63} establishes a fixed non-zero chance of US delivery every second. In particular, an RI-5 has a 20\% chance per second to deliver one drop, hence one drop falls every $1\,\textrm{s}/0.2=5\,\textrm{s}$ in average, i.e., twice per CS. Thus, an average of 88 US per session were delivered, roughly ranging between 70 and 110 drops. Each session ended 10 s after the end of the last CS.
%2

The structure of the trials for control groups was the same except for the schedule of delivery of the US: an RI-25 spanning the whole duration of the trial, so that a US could equally occur during the CS and at any other moment of the trial. This corresponds to the delivery of the same amount of solution per trial as for the experimental groups: one drop every 25 seconds in average, 4\% chance of delivery per second, average of 2 drops per trial.

The third and last phase was \emph{extinction}, the only difference with respect to previous training sessions being the absence of liquid in the drinking dispensers.

\medskip

\subsubsection{Procedure}

The subjects were distributed into four groups of 8 rats: \emph{Group 1} (subjects 1-1, 1-2, \dots, 1-8) for a US consisting in one drop of $0.1\%$ saccharin solution, \emph{Group 2} (subjects 2-1 to 2-8) for a US consisting in one drop of $0.2\%$ saccharin solution and two control groups with the same concentration (\emph{Group 1C}, subjects 1C-1 to 1C-8, 0.1\%; \emph{Group 2C}, subjects 2C-1 to 2C-8, 0.2\%) but randomized US delivery as explained in the previous subsection. Although the subjects were naïve in terms of Pavlovian explicit training, their provenance from three operant experiments was counterbalanced in each group as an extra measure of precaution to minimize the effect of uncontrollable variables due to their past history.

On the first day, all subjects went through the phase of pre-training, consisting in two consecutive sessions of the pre-training described in the previous subsection. On the same day or the day after, session 1 of the training phase was run upon the subjects. The total duration of the experiment was 90 sessions, run once or twice per day with an inter-session interval ranging from a minimum of 1 hour and 30 minutes to about 3 hours and 30 minutes. The day after session 90, we moved all subjects through two consecutive sessions of extinction.

\medskip

\subsubsection{Results}

To compare data with the theoretical model, one has to be careful about the identification of the association strength. In our experimental design, the US is presented simultaneously with the CS and one must account for the responses directly due to the US. However, conditioned and unconditioned licking are not easy to separate, since rats normally lick multiple times even for small volumes of water. Although it is difficult to eliminate this problem and thus to record pure conditioned responses, it is still possible to slightly reduce it and define the following index of learning. We can subtract the number of US (which varies from trial to trial and from session to session), to the total number of licks per session and we identify
\be\label{csus}
v=\textrm{(number of licks during CS)}-\textrm{(number of US)}\,.
\ee
Thus, whenever we talk about licks during the CS \emph{in session-by-session} data, we imply that the number of unconditioned responses (responses following the delivery of an US) has been discounted. In this way, at the beginning of training animals responded only upon presentation of the US and the total number of licks is of the same order of magnitude as the number of US presented. A minor issue is that under-response generates negative values of $v$, but this does not correspond to inhibition. While under-response at the beginning of training means that the animal does not know that the CS predicts the US, inhibition would imply that the animal knows that the CS predicts the no-US. Since negative $v$'s appear only in the very first sessions, this feature does not influence whatsoever the focus of our research on the rest of the data. 

%In Eq.\ \Eq{csus}, we defined unconditioned responses as those occurring in the presence of the US and assumed that the number of US is equivalent to the number of unconditioned responses. However, this assumption does not account for the following situation. It could be that, sometimes, the subject did not lick upon delivery of the US and that it licked instead afterwards, in the absence of the US (vacuum licking). Although a single drop stayed clung to the dispenser, the short delivery of two US could make the merger of the first and second drop fall before the animal could lick, at which point vacuum licking would take place. All these fine details are not important for the final interpretation of the results, because replacing \Eq{csus} with a different operational definition of $v$ would lead to very small quantitative differences. For example, we checked that considering, in alternative, the number of licks during the CS (not discounting the number of US), the number of licks per reinforcer (number of licks during CS)/(number of US), or the number of CS licks minus the number of post-CS licks does not change the plots qualitatively, and response variability remains at the same levels. Therefore, it is in general reasonable to assume that the associative strength is manifested in the overall behavior.

In \emph{trial-by-trial} data, we will not subtract the number of US to the actual response by the animal. The reason is that we will be interested in these data when considering the noise component of the signal (see below for details). A randomized US delivery would produce a white noise averaging to zero and of much smaller amplitude than any other source of noise (statistical error or some intrinsic effect to be checked upon).

\begin{figure*}[!ht]
\includegraphics[width=8.7cm]{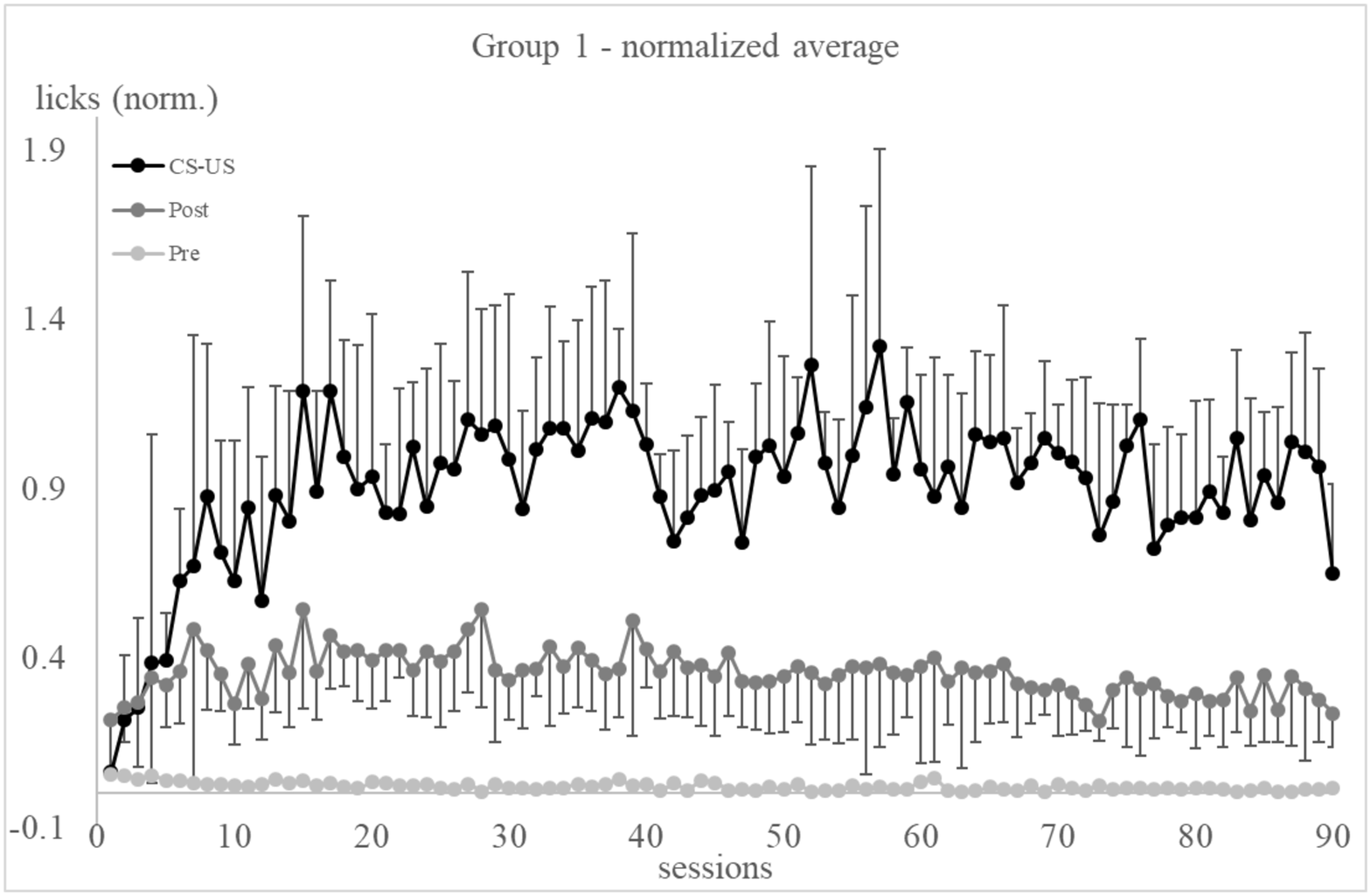}\includegraphics[width=8.7cm]{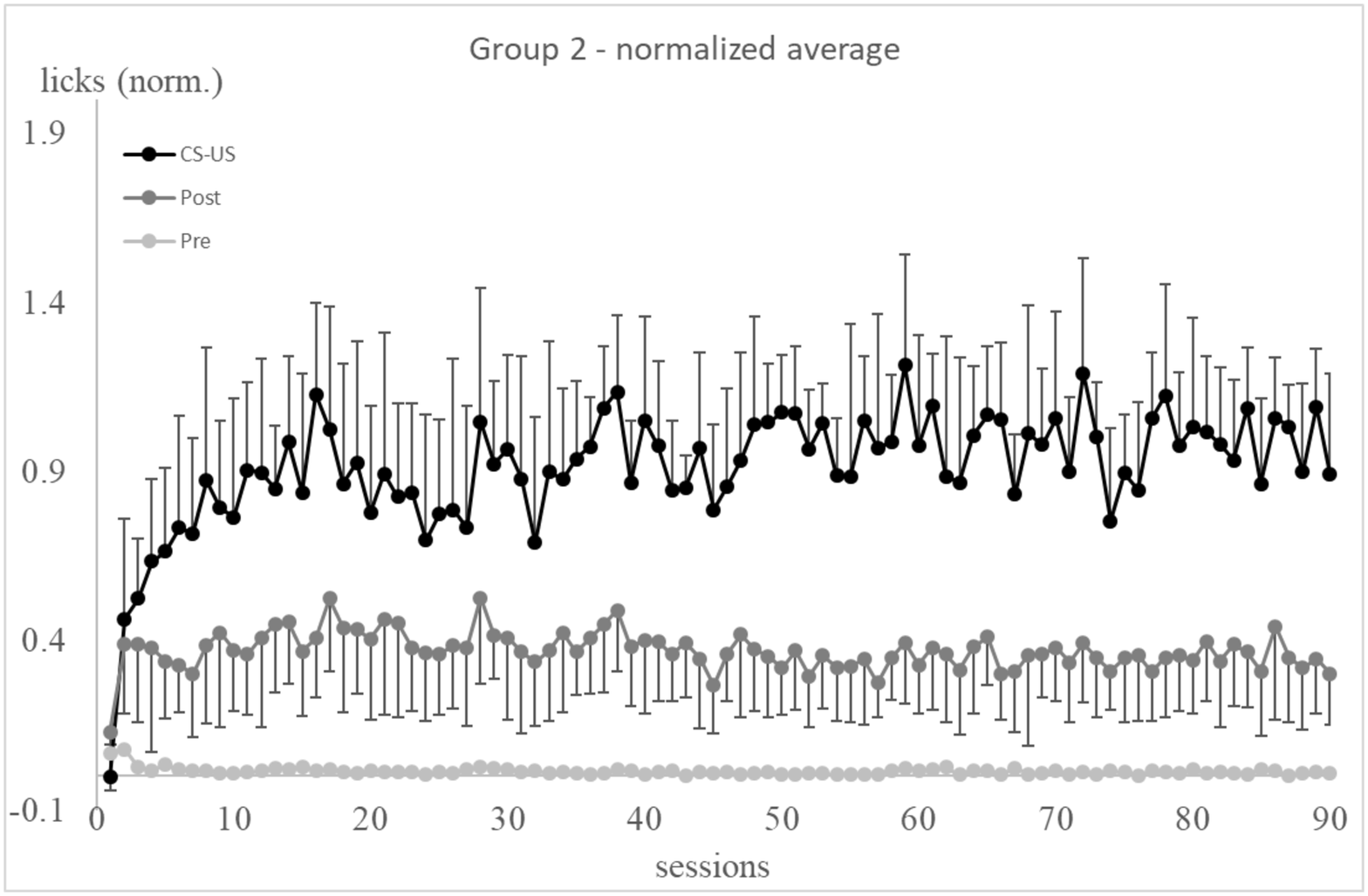}\\
\includegraphics[width=8.7cm]{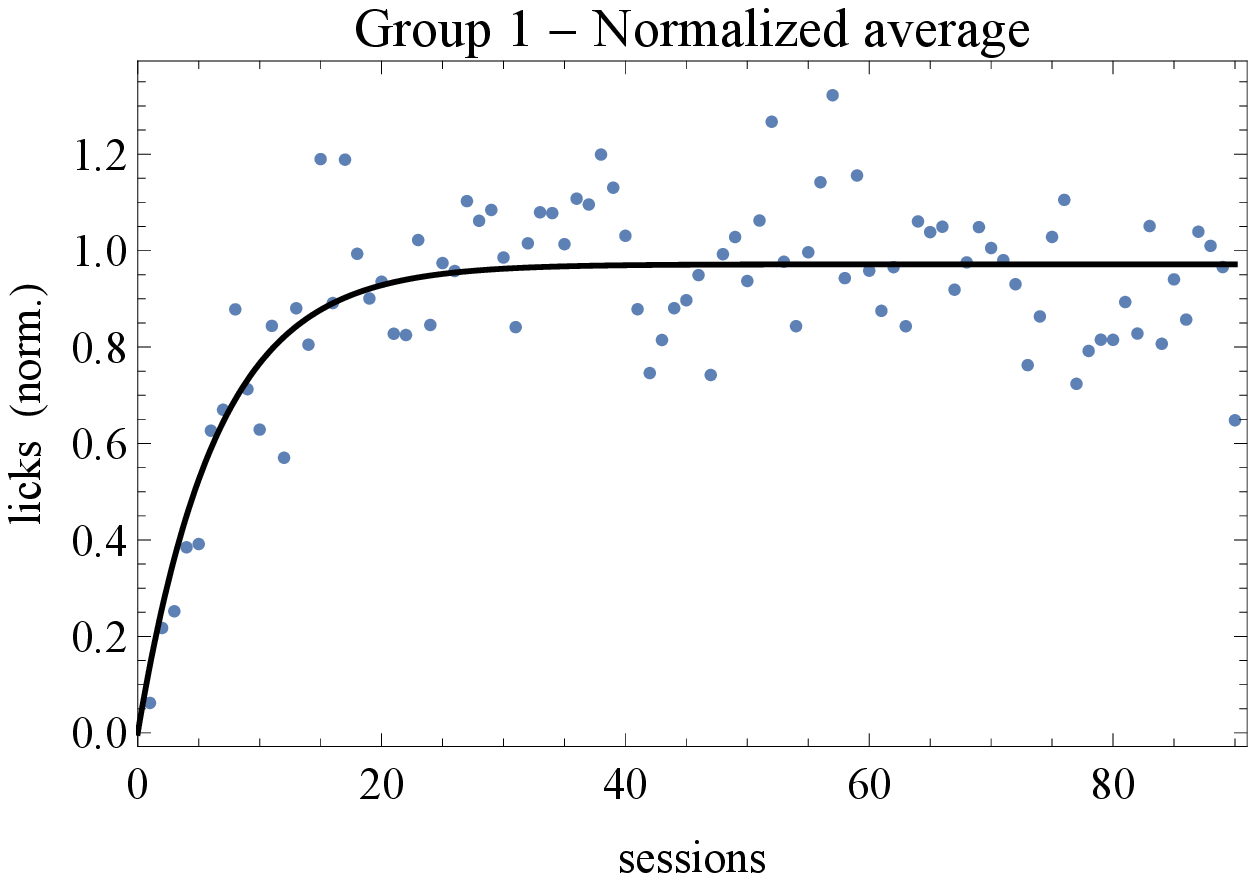}\includegraphics[width=8.7cm]{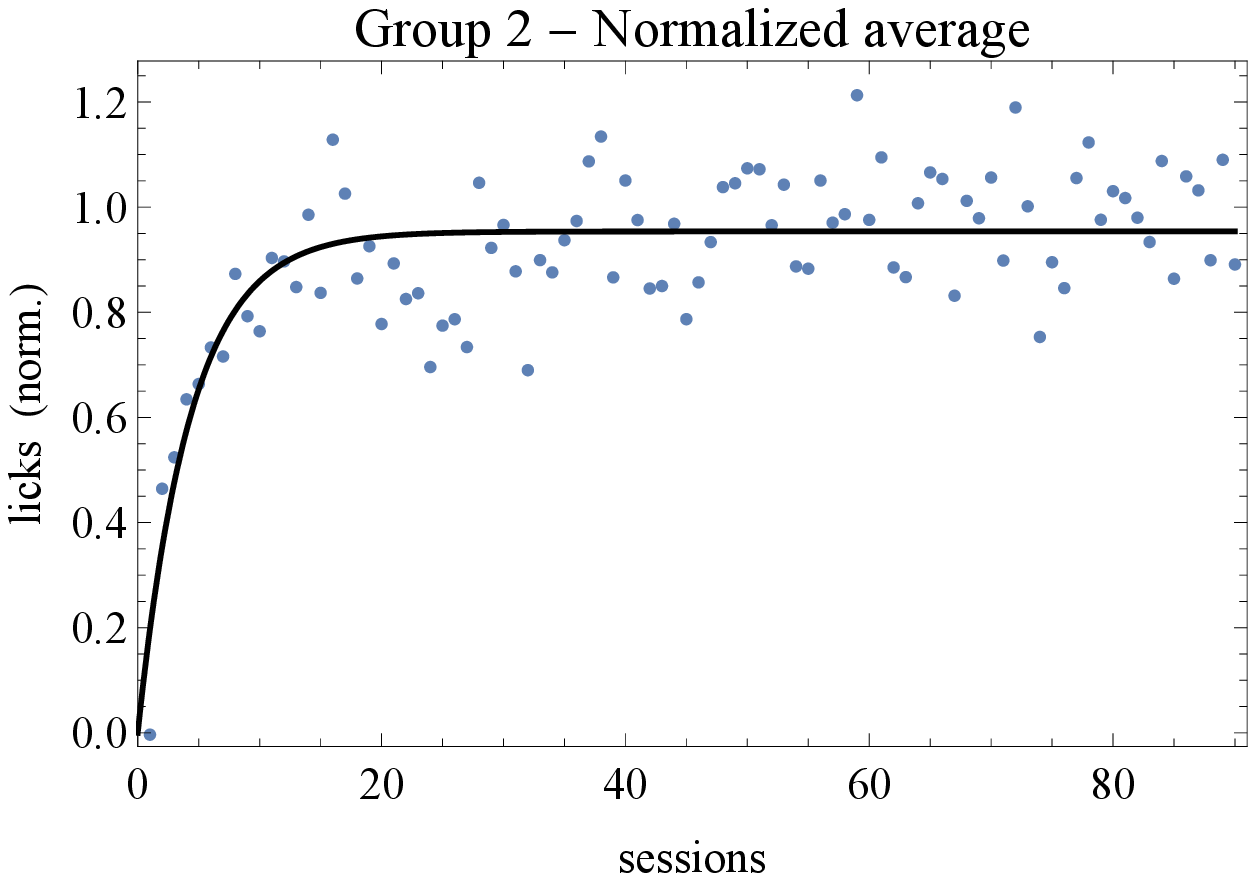}
\caption{\label{fig3} Average of normalized data for Group 1 (top left) and 2 (top right), together with the best fit with the RW model (bottom). Sessions are on the {\it horizontal axis}, number of licks on the {\it vertical axis}. Light gray, dark gray, and black data points (connected by lines of the same colors) are the licks in the 10 s respectively before, during, and after the CS. The number of US has been subtracted from the during-CS licks of each individual before taking the average, and the upper and lower error bars at the 68\% confidence level of CS and post-CS data are shown.}
\end{figure*}

After determining the asymptote of learning for each subject, we normalized the data by dividing the number of licks by the estimate of $\la$. Normalized data have some advantage over raw ones, as explained in the \textbf{Supplementary Material}. For instance, with normalized data one can directly compare the fluctuations in response of different individuals. %\footnote{Another procedure to eliminate motor individual differences not related to the learning process is to normalize each data set with respect to the maximum response {\bf [GC: referencia aquí]}. This choice does not assume the existence of an asymptote of learning and is therefore slightly model independent than ours. However, it could be argued to be too arbitrary, since it is based on the value of an \emph{a priori} random fluctuation of the response.}}
 The average of normalized data is shown in {\bf Figure \ref{fig3}}. In comparison with {\bf Figure S1}, we do not notice any qualitative change in the experimental trendline. However, error bars are considerably smaller. Also, while the estimated standard deviation $\s$ of the best fit of raw data was greater in Group 2, %(Eq.\ \Eq{Hulbf}),
 after normalization it is smaller: $\s=0.13$ for Group 1 and $\s=0.11$ for Group 2. The plots showing the data and the RW best fit of $v/\la$ for the subjects in the experimental and control groups can be found in {\bf Figures S4--S9}.
%3

Now we are in a position to comment on the results. First of all, both the pre-CS and post-CS response in the experimental groups were much smaller than the response during the CS, respectively by a factor of 10 and of 2. At the 68\% confidence level, none of the curves overlap. This means that the subjects clearly discriminated between the possibility to obtain the US during the tone and in its absence. The higher post-CS response with respect to the pre-CS is easily explained on the grounds of a natural inertia in licking that extended for a few seconds after the tone was switched off. This interpretation is confirmed by calculating Pearson's correlation coefficient for CS and post-CS licks (non-subtracted and non-normalized data) of individual subjects, which is $|r|\geq 0.45$ except in two cases (1-6 and 2-7).\footnote{This relatively high correlation might suggest that a better indicator than \Eq{csus}, stripped of micro-motivational or attentional fluctuations, could be obtained if we subtract the number of post-CS licks instead of the number of US. However, as we noted before, we verified that the results of this section do not change.} The correlation between CS and pre-CS licks is smaller, $|r|<0.38$. On the other hand, the control subjects displayed about the same response level before, during, and after the CS, indicating that there was no discrimination and no acquisition of association between the CS and the US.\footnote{The total response of control subjects in Group 1C was flat throughout the experiment: a linear fit of the total licks yields a slope $-0.23\pm 0.24$ and a correlation $r^2=0.01$. We did register a slight positive trend in Group 2C: a linear fit of the total licks yields a slope $1.43\pm 0.31$ and a correlation $r^2=0.20$. This trend was driven by subjects 2C-2, 2C-5, and 2C-8. Possibly, this means that subjects developed a liking (unrelated to any CS-US associative process) for the saccharin solution at 0.2\% more pronounced that for the less concentrated solution.} Moreover, the average asymptotes of learning of Groups 1 and 2 reported in the \textbf{Supplementary Material} %Eq.\ \Eq{Hulbf}
 before normalizing the data are significantly different: the rats did respond differentially to 0.1\% and 0.2\% concentrations. Finally, the data of extinction show a very quick decrease in response of experimental subjects, likely due to overtraining \citep{Fin42}. The flat response of control subjects during the CS ({\bf Figure S2}) reflects the absence of association. Also, control subjects showed extinction to the whole ``apparatus + tone + electrovalve click'' stimulus (light gray trendlines). Overall, these results indicate that simultaneous conditioning was effective and they validate the experimental design.

Let us now turn to individual response differences. Without the pretense of being exhaustive, we registered the following two pairs of patterns:
\bi
\item \emph{Wildly fluctuating response} (e.g., subjects 1-1, 1-4, 1-7, 2-1, 2-2, 2-5, 2-6, 1C-1, 1C-3, 1C-6, 2C-1, 2C-4). For experimental subjects, the shape of the learning curve is almost completely lost into noise.
\item \emph{Less fluctuating response} (e.g., subjects 1-2, 1-3, 1-5, 1-6, 2-3, 2-4, 2-7, 2-8, 1C-2, 1C-4, 1C-7, 1C-8, 2C-2, 2C-3), where strong fluctuations are less frequent. Experimental subjects follow more clearly the learning curve.
\ei
The difference between ``wild'' and ``mild'' fluctuations can be quantified by considering the estimated standard deviation of data with respect to the nonlinear fit with the RW learning curve ({\bf Table \ref{tab1}}). We took as a criterion for ``wild fluctuations'' variations greater than or equal to 30\% of the peak response ($\s\geq 0.30$). This criterion is arbitrary (one could have taken, e.g., $\s=0.50$ as a threshold) but illustrates the point. Some subjects are somewhat in between these two categories, since they showed a stable response during a long time followed by wildly fluctuating periods.
%1
\begin{table}
\begin{center}
\begin{tabular}{cccccccc}\hline
												\textbf{1-1} &\textbf{1-2} & \textbf{1-3} & \textbf{1-4} & \textbf{1-5} & \textbf{1-6} & \textbf{1-7} & \textbf{1-8}\\\hline
 0.59 & 0.29 & 0.19 & 0.33 & 0.16 & 0.24 & 0.43 & \\\hline
\textbf{2-1} & \textbf{2-2} & \textbf{2-3} & \textbf{2-4} & \textbf{2-5} & \textbf{2-6} & \textbf{2-7} & \textbf{2-8}\\\hline
 0.32 & 0.40 & 0.22 & 0.20 & 0.36 & 0.30 & 0.10 & 0.15 \\\hline
\end{tabular}
\caption{\label{tab1} Estimated standard deviation $\s$ of the RW-model best fit of normalized session-by-session data.}
\end{center}
\end{table}

For those subjects that showed a trend in their response, we can further recognize:
\bi
\item \emph{Slow increase in response} (e.g., subjects 1-2, 2-7, 1C-6, 2C-2, 2C-5, 2C-8). For experimental subjects, this is simply due to a slow learning rate, while the interpretation for control subjects is less obvious. Perhaps the unpredictability of the US was a factor increasing the response, as observed also by \citet{KP84}.
\item \emph{Slow decrease in response} (e.g., subjects 1-1, 1-7, 1C-1, 1C-7, 2C-7). This phenomenon is related to the presentation of the stimuli and their mutual association. This is not short-term habituation, which is a non-associative process occurring relatively quickly (only a few sessions) \citep{Cev14,Tho09}. Moreover, short-term habituation is faster for weaker stimuli, which we did not see here. However, it is not long-term habituation either, which also occurs in a relatively short time span (a few sessions, although it depends on the experiment and the subject species) before any plateau in the response \citep{OrGu,PaSi,Pla97}. Here one might rather talk about a very long-term habituation, occurring because animals are presented with the same stimulus in all trials of all sessions.
\ei
In the next sections, we will make a more in-depth analysis of the individuals' data.

\medskip

\subsubsection{Discussion}

Summarizing the conclusions obtained from the above results:
\bi
\item The experimental design has been validated as a viable tool of generation and control of Pavlovian conditioning.\footnote{It is inevitable that the procurement of saccharin in the present preparation implies licking at the bottle spout, thus establishing an operant contingency between licking and obtaining the reinforcer. Furthermore, given that the US occurred at random times within the CS, this is an ideal condition for the maintenance of superstitious licking. Despite this being correct \citep{KP13,PIK,PK15}, lick suppression has been accepted as Pavlovian under an experimental paradigm similar to the current one, when the consequence is aversive or has been devalued (e.g., \citealp{JWM}). By analogy, lick enhancement might have those same characteristics as the reduction of the response, not to mention that after extended training it has been generally accepted that behavior shifts control from the consequence to the antecedent stimulus \citep{DiBa}, a case in which the results that will be modeled here might fall. Additionally, even if we accept that licking in the present experiment has an operant contribution to its installment, which is true, we believe that the theoretical analysis applied in the present paper is not really affected if an event stimulus is replaced by a response (see also section \ref{geco} \emph{Conclusions}).} The subjects discriminated between the different chances of getting the US during the tone CS (simultaneous conditioning) with respect to when the tone was absent. The response was much higher during the CS than before or after. The definitory criterion for successful discrimination is the ratio between pre-CS and CS licking. Post-CS response, seldom discussed in the literature, was much larger than pre-CS response due to a natural inertia in the licking behavior but, still, it was much smaller than the response during the CS.
\item While taking the average of raw data is useful for between-groups comparisons (absolute value of the asymptotes), error bars are reduced when considering the average of data with normalized asymptote of learning. Normalizing also the learning rate of individuals does not add much information and is a strongly model-dependent procedure.
\item Although we did observe long-range (i.e, spanning several sessions) fluctuations in the average subjects response, the error bars due to individual differences are large enough to conclude that these fluctuations are not significant. The RW model \Eq{DeV2av} is a good description of the average learning curve in Pavlovian conditioning.
\item The main average effects of overtraining are an extremely slow decrease of the post-CS response and a fast extinction (see {\bf Figure S10}). Fast extinction points out that the response of the animals during the experiment was not driven by habit, contrary to what one might expect in long training histories \citep{GDB}. %Although we cannot offer the reader quantitative data, we could also observe a very fast recuperation of the response in an informal post-extinction session for randomly chosen experimental and control subjects {\bf [GC: aquí lo que pasó en la realidad es que se observó recuperación después de sesiones malas donde no habían caído gotas.]} and the gradual development of slight hyperactivity in control subject.
\ei

We can compare our findings with those in the literature of post-peak depression or inhibition by reinforcement cited in the \emph{Introduction}. In general, our subjects reached the asymptote of learning after 15 to 20 sessions. While gradients in response have been registered on as short a scale as trial-by-trial or session-by-session intervals, large-scale (i.e., spanning many sessions) fluctuations characterized all the plateau after acquisition. In some cases, we did see a response decrease, but much later than acquisition. Even granting that aversive conditioning may be faster than appetitive one, this leads us to believe that this decrease is a long-range phenomenon different from the post-peak depression observed in experiments employing only a few hundred CS-US pairings, in contrast with our almost four thousand trials each with an average of two CS-US pairings. The latter could be a transient phenomenon corresponding to the first fluctuation peak just after acquisition, when present. Such interpretation is corroborated by past evidence on the non-robustness of inhibition by reinforcement when extending the duration of the experiment \citep{DuLy}. 

In general, observations of individual differences were not accompanied by attempts to explain them quantitatively (see, however, \citealp{UWM}). In the following sections we want to do just that. It should be made absolutely clear that the fact that the RW model is a good fit of data does \emph{not} mean that individual differences and response fluctuations are mere statistical phenomena to be treated as unwanted errors. Different subjects do respond very differently to stimuli and their response does change erratically trial after trial and session after session. The issue then is whether we can find a theoretically motivated model (not just an \emph{ad hoc} fitting curve, which is not hard to concoct) better than the RW model in explaining the data, in particular, the long-range response decrease observed in some subjects (not to be confused with the post-peak depression effect in the literature, as already said above).

%%%%%%%%%%%%%%%%%%%%%%%%%%%%%%%%%%%%%%%%%%%%%%%%%%%%%%%%%%%%%%%%%%%%%%%%%%%%%%

\medskip

\subsection{Experiment 2 of Harris et al.\ (2015)}\label{two2}

This experiment was described in detail in \citet{HPG} under the heading ``Experiment 2.'' Here we will briefly review its characteristics. Data will be analyzed in the next section.

\medskip

\subsubsection{Subjects and materials}

Subjects: 16 experimentally naive male Hooded Wistar rats, 8 to 10 weeks of age at the start of the experiment. They had unrestricted access to water and restricted daily food rations of regular dry chow equal to 5\% of the total weight of all rats in the tub, provided 30 minutes after the end of the daily training session.

Materials: 16 Med Associates conditioning chambers with a food magazine endowed with an infrared LED and sensor inside, to record entries by the rat. A dispenser delivering 45 mg food pellets was attached to the food magazine. Four different CSs were used: white noise (78 dB), a tone (78 dB, 2.9 kHz), a flashing light (2 Hz, 3.0 cd/m$^2$), and a steady light (30 cd/m$^2$).

\medskip

\subsubsection{Experimental design and procedure}

\begin{figure}[!ht]
\includegraphics[width=8.5cm]{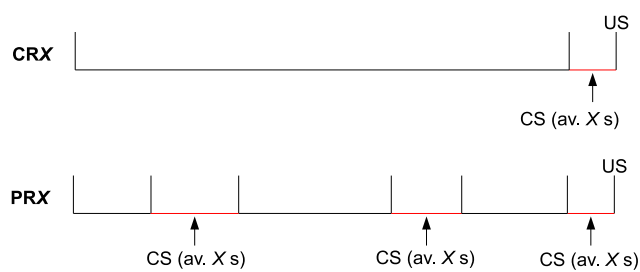}
\caption{\label{fig4} Structure of a trial for the four experimental groups of Experiment 2 of \cite{HPG}, where $X=10$ or 30.}
\end{figure}

For each rat, each CS was allocated to one of the following configurations (and counterbalanced across rats):
\bi
\item CR10: continuous reinforcement (CS reinforced 100\% of the times) with random duration of 10 s mean. 30 sessions, 6 CR10 trials per session.
\item CR30: continuous reinforcement with random duration of 30 s mean. 30 sessions, 6 CR30 trials per session.
\item PR10: partial reinforcement (CS reinforced at 33\% of the times) with random duration of 10 s mean. 30 sessions, 18 PR10 trials per session.
\item PR30: partial reinforcement with random duration of 30 s mean. 30 sessions, 18 PR30 trials per session.
\ei
Trial by trial, the CS duration varied randomly on a uniform distribution with a mean of either 10 s (2 to 18 s) or 30 s (2 to 58 s).
The number of reinforced trials per session per CS was the same in all configurations and equal to 6. See {\bf Figure \ref{fig4}} for a scheme of the trial structure of this experiment.
%4

Each of the 30 sessions consisted in a delayed conditioning where presentations of each of the four CSs were randomly intermixed: 6 of
each of the continuously reinforced CSs and 18 of each of the partially reinforced CSs, for a total of 48 trials per session.
The ITI varied randomly on a uniform distribution with a minimum of 50 s and a mean of 120 s. 

The subject response was the head entry into the magazine, recorded during each CS and a 30-second pre-CS period through the interruption of a photobeam at the magazine entrance.

\medskip

\subsubsection{Results}

We analyzed the data considering each CS configuration as completely independent of the others. Thus, we treated CR10, CR30, PR10, and PR30 as four independent groups, each undergoing 30 sessions of, respectively, 6, 6, 18, and 18 trials, for a total of, respectively, 180, 180, 540, and 540 trials.

Sessions 26-30 had 7 trials each for the CR10 and CR30 CSs, rather than 6 trials. The reason of the inclusion of the extra trial (which was without reinforcement) was to look at the post-CS response for at least one trial of the two CR CSs that was not contaminated by the presence of the food-pellet US. Although their impact on the results is negligible, we removed these 5 extra trials from the analysis.

Also, a word of caution on the analysis of the PR groups. PR subjects underwent partial reinforcement, which means that theoretical models based on continuous reinforcement (for instance, plain excitatory Rescorla--Wagner) should be applied with care, since they do not take into account the fact that the US was not presented at all trials. However, when the presentation rate is deterministic, fitting PR data entries with models of deterministically delivered continuous reinforcement will be perfectly consistent.

%%%%%%%%%%%%%%%%%%%%%%%%%%%%%%%%%%%%%%%%%%%%%%%%%%%%%%%%%%%%%%%%%%%%%%%%%%%%%%
%%%%%%%%%%%%%%%%%%%%%%%%%%%%%%%%%%%%%%%%%%%%%%%%%%%%%%%%%%%%%%%%%%%%%%%%%%%%%%

\section{Dynamical model of individual behavior}\label{dyns}%\section{DYNAMICAL MODELS OF INDIVIDUAL BEHAVIOR}

%%%%%%%%%%%%%%%%%%%%%%%%%%%%%%%%%%%%%%%%%%%%%%%%%%%%%%%%%%%%%%%%%%%%%%%%%%%%%%

\medskip

\subsection{Motivation: least-action principle and fine tuning}

Thanks to its simplicity, the RW model is an ideal example where to introduce all the main ingredients of a dynamical reinterpretation of Pavlovian conditioning processes. By \emph{dynamical}, we mean a very precise concept, superior to any casual use of the term in the loose sense of ``evolving'' or ``interacting''. Namely, we postulate that any conditioning process can be described by a quantity called \emph{action} and that the change of the association strength during conditioning happens in such a way that the action is minimized. Let us introduce the rationale behind this view.

As a global, externally observable phenomenon, Pavlovian learning has been described through models such as Rescorla--Wagner's \citeyearpar{RW72} and others \citep{LeP,Mac75,PH,Wag81,WV}. In general, the essence underlying these models appeals to psychological aspects such as the surprisingness or novelty or predictiveness of the stimuli. However, it is not unreasonable to believe that an alternative conceptualization is possible where, despite behavioral errors by the subject, the learning process is a naturally efficient one and, so to speak, minimizes the biological adaptation effort. At a biological level, learning can be viewed as a sequence of events modifying some of the synaptic connections of the brain. This modification does not happen in a disordered way since, as associative models already highlighted, there exist laws (learning curves) applicable to statistically significant samples. Going a bit beyond the macroscopic view of traditional associative models, but without attempting a microscopic quantitative description of neural plasticity, we postulate that the brain mechanics of a subject change through learning from an initial state A to a final state B efficiently. Operationally, a most effective way to describe this minimization of effort is through the action. If we depict the learning process in time $t$ as a path from A to B in an abstract space parametrized by the association strength $v$, the profile $v(t)$ describing the evolution in the association strength minimizes the path from point $v(t_{\rm A})$ to point $v(t_{\rm B})$. A similar statement could be made about the energy spent in changing the internal state, but both are described by the same quantity, the action. The action $S[v]$ is a function of the association strength $v$ and it is minimized when its variation with respect to $v$ is zero:\footnote{At the maxima and minima of a function, its first derivative vanishes.}
\be\label{pla}
\frac{\de S[v]}{\de v}=0\,.
\ee
This is the principle of least action. It has been applied successfully in physics and our aim now is to use it also in psychology. From physics we will get guidance about what $S[v]$ is and this guidance will prove itself correct because it will immediately recover the RW model as a special case.

\begin{figure}[ht!]
\includegraphics[width=8.5cm]{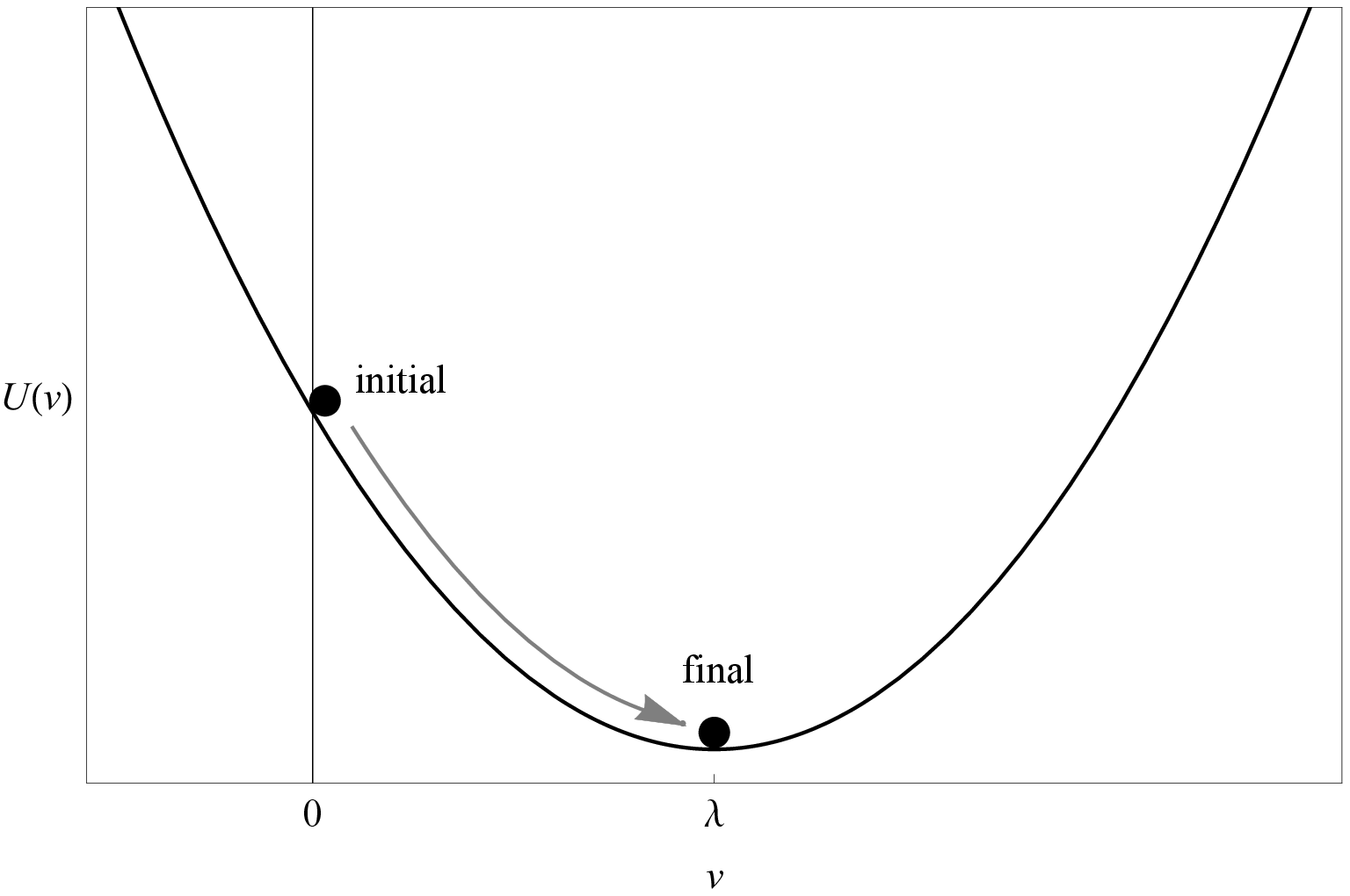}\\
\includegraphics[width=8.5cm]{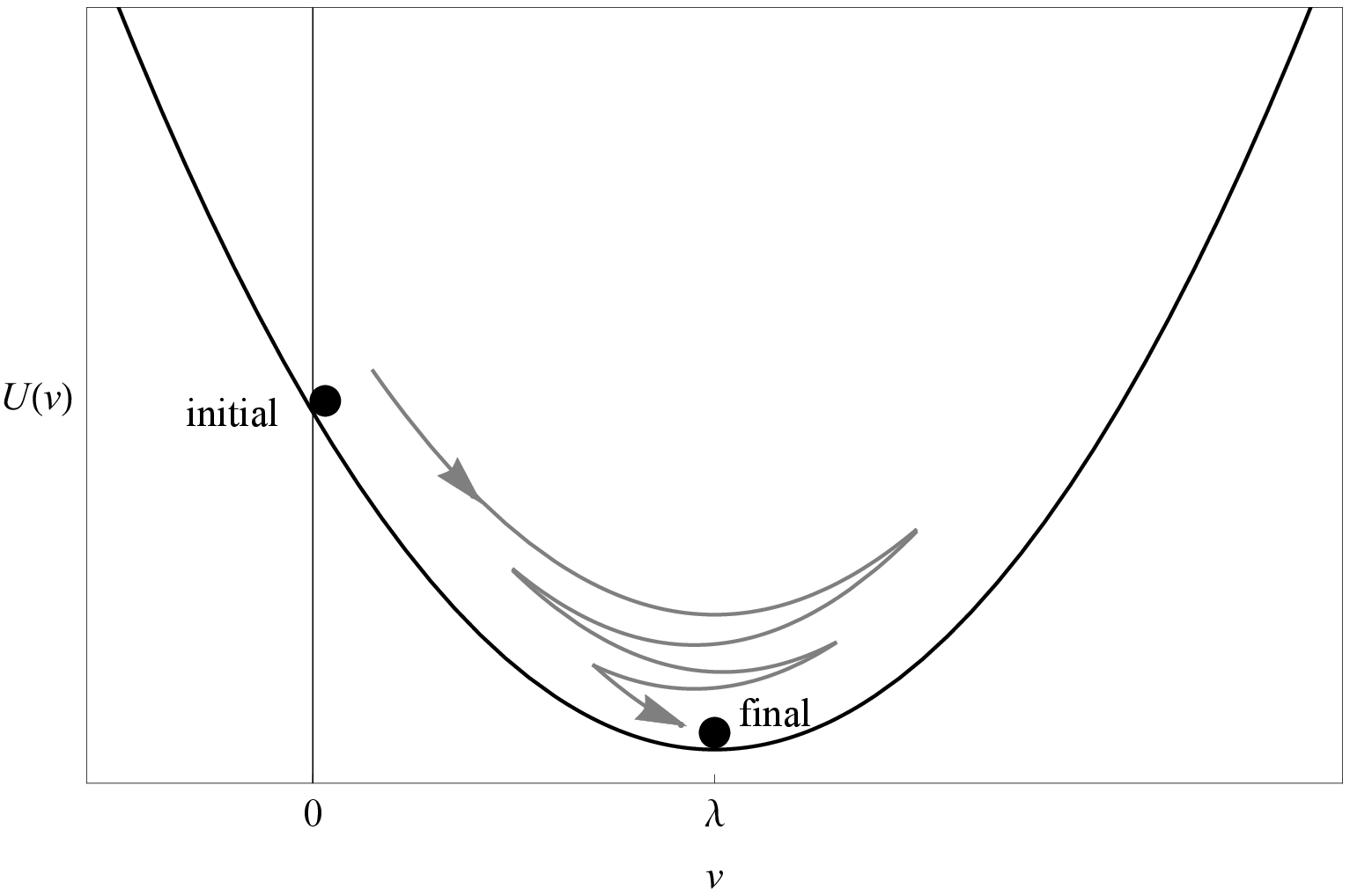}
\caption{\label{fig5} {\bf (Top)} The potential $U(v)$ of the RW model %\Eq{S2b}
 for $\la=1$ and $\a\b=1$. Compare with the solution \Eq{sol1} in {\bf Figure \ref{fig1}}. {\bf (Bottom)} The potential $U(v)$ of the dynamical model %\Eq{pote2}
 for $\la=1$ and $(\a\b)^2+\mu^2=1$. Compare with the solution \Eq{oscso} in {\bf Figure \ref{fig6}}. In both cases, the particle rolling down the potential represents the change in the associative strength. The direction of ``motion'' in excitatory conditioning is represented by a gray arrow.}
\end{figure}

The main idea is to reinterpret the learning curve of any associative model as the trajectory of one or more small balls (pointwise \emph{particles}) rolling up and down a hill (a \emph{potential}). The way a particle moves along its potential is called dynamics. In the case of the RW model, there is only one particle whose trajectory is shown in {\bf Figure \ref{fig1}} and whose potential $U(v)$ is depicted in {\bf Figure \ref{fig5}} (top). The proof of this statement, the explicit form of the action $S[v]$ and the dynamics corresponding to multi-cue or variable-salience conditioning models are given in the \textbf{Supplementary Material}. In excitatory conditioning with just one cue, the particle rolls down the slope from the point $v=0$ to the bottom at $v=\la$, where it unnaturally stops.
%5

At this point, one appreciates the first major advantage of the least-action principle. If the latter is true, then it is very hard to understand why a particle placed on the slope of the potential well would roll down and stop exactly at its bottom. The least-action principle tells us that we must fine tune the initial conditions (position and velocity) of the particle to infinite precision in order to achieve such a behavior. If some latitude in the choice of initial conditions is allowed (as it should in a natural biological setting), then in general the particle will oscillate up and down the well until reaching a final stop at the bottom, as shown in the bottom plot of {\bf Figure \ref{fig5}}.

We decided to use tools borrowed from physics because, in the long run, they carry two major advantages. First, as said above, they allow one to modify Pavlovian models in a natural way that would result rather obscure in the traditional approach, and that can be contrasted with experiments. Second, they are the basis from which one can construct predictive theories of individual short-scale response variability, presented in the \textbf{Supplementary Material}. Ultimately, revisiting conditioning models as dynamical models amounts to a new paradigm of doing model-building in psychology, where qualitative reasonings leading to quantitative formul\ae\ are replaced by a rigorous sequence of logical steps. As in any model building, arbitrariness is not removed, but it will be pinpointed and put under a higher degree of control.

%%%%%%%%%%%%%%%%%%%%%%%%%%%%%%%%%%%%%%%%%%%%%%%%%%%%%%%%%%%%%%%%%%%%%%%%%%%%%%

\medskip

\subsection{Theory}\label{secthe}

\begin{figure}[!ht]
\includegraphics[width=8.5cm]{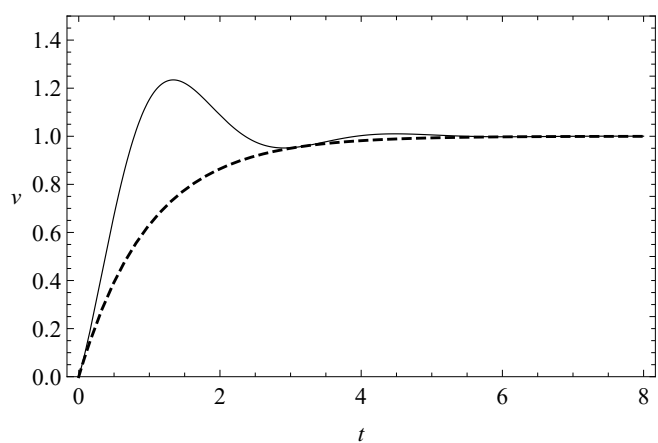}
\caption{\label{fig6} The learning curve \Eq{oscso} (solid) with periodic oscillations with $\mu=2$ and $A=0$, compared with the RW learning curve \Eq{Hullearn} (dashed). Here $\la=1=\a\b$.}
\end{figure}

The catchword is ``unnatural.'' The way the particle moves along its potential in the case of the RW model is very special because it reduces to the simple equation \Eq{DeV2}. When it rolls down the slope, the particle experiences some resistance (\emph{friction}) from the floor, but not so much as to brake completely. This happens, by sheer coincidence, exactly at the bottom of the slope. In a more general situation, we would expect the particle to oscillate up and down the bottom, if the friction is moderate, until it reaches a complete stop ({\bf Figure \ref{fig5}}, bottom). If we abandon the rigid setting of the RW dynamics and allow for such a scenario, much more natural from a dynamical point of view, we obtain a different trajectory, i.e., a different learning curve (see \textbf{Supplementary Material}):
\be\label{oscso}
v(t)=\la\left[1-\rme^{-\a\b t}(\cos\mu t+A\sin\mu t)\right],
\ee
where $\mu$ and $A$ are constants. When $\mu=0$, there are no oscillations and this reduces to the RW model \Eq{Hullearn} of excitatory conditioning. When $\mu\neq 0$, learning is subject to a friction force with progressively damped oscillations around zero. The parameter $A$ is the amplitude of the sine harmonics, so that when $A=0$ oscillations have a pure cosine phase. The learning curve is modified as in {\bf Figure \ref{fig6}}. This profile features oscillations of fixed frequency $\mu$ and decreasing amplitude above and below the asymptote, which we can look for in data. In particular, it predicts a first response peak above the subsequent asymptote.
%6

Notice that $v(t)$ is not positive definite at small times unless $A$ is sufficiently small. To avoid this problem and to remove a free parameter from the model, we will pay special attention to the case $A=0$. Data will show that the $A\neq 0$ is disfavored anyway.

Reverting back to discrete time, we can recast the DOM %Eq.\ \Eq{eomH2}
 as a law for the association strength $v_n$ at the beginning of trial (or session) $n$. The first and second-order derivatives $\dot v$ and $\ddot v$ correspond, respectively, to the forward finite differences $\De v_n=v_n-v_{n-1}$ and $\De v_{n+1}-\De v_n=v_{n+1}-2v_n+v_{n-1}$, so that the analogue of Eq.\ \Eq{DeV} is 
\be\label{OMdis}
\De v_{n+1}-(1-2\a\b) \De v_n=(\a^2\b^2+\mu^2)(\la-v_{n-1})\,.
\ee
It is easy to show that the RW model is recovered when $\mu=0$. In fact, \Eq{DeV} implies that $\De v_{n+1}=\a\b(\la-v_n)=\a\b(\la-2\De v_n+ \De v_n-v_{n-1})=-2\a\b\De v_n+\a\b(\la-v_{n-1})+\a\De v_n=(1-2\a\b)\De v_n+(\a\b)^2(\la-v_{n-1})$, which is Eq.\ \Eq{OMdis} in the limit $\mu\to 0$. Unless $\mu=0$, it is not possible to write a simple incremental law as RW, and the association strength at the beginning of trial $n+1$ is predicted by that of the previous \emph{two} trials $n$ and $n-1$, instead of just the one immediately preceding as in the RW model \Eq{DeV}. Both in the RW model and in the DOM one can predict the value $v_{n+1}$ at the next trial, but while in the RW model we only need to know the present value $v_n$, in the DOM we also need the past value $v_{n-1}$. Shifting this statement one trial in the past is equivalent to say that the DOM can predict the value of the association strength not only at the next trial, but also at the next-to-next one, provided we know the present value. Conversely, shifting the last statements two trials in the future, from the present and the previous value of the association strength the DOM can retrodict the value of the association strength two trials before. In this sense, the DOM has a longer memory of past states than the RW model.

Since this feature gives valuable insight in the psychological interpretation of the DOM, let us dwell more on it. In the RW model, the main postulate is that learning (i.e., the momentum $\De v_n$, the difference of association strength between one trial and the previous) is proportional to the prediction error $(\la-v_{n-1})$. This is the incremental law \Eq{DeV}. In the 
RW model, the increment $\De v_n$ is always positive, the association strength increases and the subject always ``learns more'' than before. However, in the DOM the association strength of the subject can also decrease in some trials, corresponding to the descending slope in the learning curve after a peak. This does not mean that the subject has ``unlearned'' what was previously acquired; this would be true if learning were a positive monotonic process as in the RW model. Rather, the association strength decreases only after having increased to an over-optimal point. The subjects are neither over-learning nor unlearning: they are simply trying to reach a balance between what was learned and their capacity to predict the environment.

This is reflected in the equation governing the DOM, which is more complicated than the incremental law \Eq{DeV} and is replaced by Eq.~\Eq{OMdis}. This expression relates knowledge, predictions and retrodictions of three different trials in terms of the prediction error. It is as if the animal had a better predictive algorithm than in RW in terms of how many future and past trials it can scan. In particular, Eq.~\Eq{OMdis} implies that the prediction error depends mainly on the learning at the next-to-next trial and, to a lesser degree (because $1-2\a\b<1$), on the learning at the next trial. However, with respect to RW prediction errors weigh more on learning, since the right-hand side of \Eq{OMdis} is augmented by a factor $\mu^2$. Hence subjects can predict what is going to happen further in the future, but they do so with greater uncertainty. Or, alternatively, prediction of the subject performance at a given trial requires more information from the past, which introduces a heavier weight of the error. This is the origin of the overshooting of optimal response and the subsequent readjustments through an oscillatory pattern. We dub this justification of the DOM the \emph{errors-in-learning argument} and it is perhaps the most compelling one from a psychological point of view. It makes the learning process considerably more flexible than in the RW model.

The description of learning with second-order differential or finite-difference equations (momenta) has a precedent in connectionism \citep{Rum86}, while a parallel to frictional kinematics was also established by \citet{Qia99}. As far as we know, these ideas have never been applied to animal learning. Also, our theoretical motivations are different and strongly based on a dynamical (action-based) view of learning and behavioral processes, rather than on a simple analogy with kinematical equations.

%%%%%%%%%%%%%%%%%%%%%%%%%%%%%%%%%%%%%%%%%%%%%%%%%%%%%%%%%%%%%%%%%%%%%%%%%%%%%%

\medskip

\subsection{Data analysis: comparison with the RW model}

In order to assess the goodness of fit of the dynamical proposal with respect to the traditional RW model, we should take into account that the first has more free parameter than RW. Having more parameters clearly gives an intrinsic flexibility in data fitting that more rigid theories do not have, and one should balance this factor against the actual capabilities of the new models to accommodate observations. This is the typical situation where comparative statistics such as the Bayesian (or Schwarz) Information Criterion (BIC) and the Akaike Information Criterion (AIC) can be extremely useful, as advocated by \cite{WHM}.

Let $\s_{\rm e}^2:=\sum_{n=1}^N [y_n-f(t_n)]^2/N$ be the error variance of the fit, the averaged sum of squared residuals (also known as residual sum of squares or sum of squared errors), where $N$ is the number of data, $y_n$ is the experimental datum at the $n$-th session and $f(t_n)$ is the value predicted by the theoretical model. (The error variance $\s_{\rm e}^2$ is not the estimated variance $\s^2$ of the fits discussed above. They are related to each other by $\s_{\rm e}^2=N\s^2/(N-p)$, where $p$ is the number of free parameters of the function $f$.) The BIC is defined as \citep{Sch78}
\be\label{bic}
\textrm{BIC}:= N\ln\s_{\rm e}^2+p\ln N\,,
\ee
while the AIC is \citep{Aka74}
\be\label{aic}
\textrm{AIC}:= N\ln\s_{\rm e}^2+2p\,.
\ee
The first term in Eqs.\ \Eq{bic} and \Eq{aic} quantifies the badness of the fit (the greater the error variance, the worse the fit), while the second term increase linearly with $p$. The BIC and AIC penalize model complexity slightly differently. For each theoretical model, one can compute the BIC and the AIC: the model with smaller criteria is to be preferred. Calling the difference $\De:=|\textrm{(IC model 1)}-\textrm{(IC model 2)}|$ for the Bayes or Akaike IC, one finds weak evidence if $\De<2$, positive evidence if $2\leq\De<6$, strong evidence if $6\leq\De<10$, and very strong evidence if $\De>10$ \citep{Jef61,KR95}. We will use the term ``moderate'' to indicate cases where evidence is positive in one criterion but weak in the other.

While the RW model has two free parameters ($\la$ and $\a\b$), the DOM \Eq{oscso} has four ($\la$, $\a\b$, $\mu$, and $A$), which can give more flexibility with respect to the RW model but are more penalized in the Information Criteria when fitting session-by-session individual data.

We can divide the subjects in three groups: those for which the RW model is clearly favoured, those for which the DOM is clearly favored, and those where the RW and dynamical oscillatory models are about equally favored by the Information Criteria.

\medskip

\subsubsection{Our experiment}

The results for our long experiment are summarized in {\bf Tables \ref{tab2}}, {\bf \ref{tab3}}.
%2
\begin{table}
\begin{center}
%\begin{tabular}{|c|cccccccc|}\hline
\begin{tabular}{lcc}\hline
{\bf Subject} &	{\bf RW}    & {\bf Oscillations} \\\hline% & fractional
1-1& {\bf 171}, {\bf 163} & $175$, $165$ \\%& 175, 165 \\
1-2& {\bf 43}, {\bf 36} & $48$, $38$ \\% & 80, 70 \\
1-3& {\bf $-$32}, {\bf $-$40} & $-28$, $-38$ \\% & $-28$, $-38$ \\
1-4& {\bf 68}, {\bf 60} & $71$, $61 $\\% & 62, 52 \\
1-5& {\bf $-$62}, {\bf $-$69} & $-57$, $-67$ \\% &  $-44$, $-54$ \\
1-6& {\bf 13}, {\bf 5} & $18$, $5$ \\% & 41, 31 \\
1-7& $116$, $108$ & {\bf 91}, {\bf 78} \\% &  123, 113 \\
1-8&-- & -- \\%\hline% & -- \\\hline
2-1& {\bf 64}, $56$ & $65$, {\bf 55} \\% & 68, 58 \\
2-2& {\bf 103}, {\bf 95} & $107$, $97$ \\% & 110, 100 \\
2-3& $-3$, $-10$ & {\bf $-$4}, {\bf $-$14} \\% & 5, $-5$ \\
2-4& $-27$, $-35$ & {\bf $-$33}, {\bf $-$43} \\% & $-16$, $-26$ \\
2-5& {\bf 84}, $76$ & $85$, {\bf 75} \\% & 88, 78 \\
2-6& {\bf 49}, {\bf 41} & $52$, $42$ \\% & 52, 42 \\
2-7& {\bf $-$139}, {\bf $-$147} & $-135$, $-145$ \\% & $-64$, $-74$ \\
2-8& $-75$, $-82$ & {\bf $-$76}, {\bf $-$86} \\\hline% & $-70$, $-80$\\\hline
\end{tabular}
\caption{\label{tab2} BIC and AIC (in the format ``BIC, AIC'' and approximated to zero decimals) of the best fits of the session-by-session data of our long experiment with the RW model \Eq{Hullearn} and the DOM \Eq{oscso}. %, and the fractional model described in the \textbf{Supplementary Material}.
 Favored models are in boldface. %, except the fractional model for subject 1-4 because it is a false positive.
 Although we fitted the DOM in two ways, one with $A=0$ from the start and one with $A$ free, we report only the best between these two versions.}
\end{center}
\end{table}
%3
\begin{table}
\begin{center}
\begin{tabular}{lccc}\hline
     & {\bf Group (\%)} 1 & {\bf Group 2 (\%)} & {\bf Total (\%)} \\\hline
RW & 86 & 50 &  67\\
Oscillatory & 14 & 50 & 33\\\hline
\end{tabular}
\caption{\label{tab3} Percentage of subjects of our long experiment following the RW model or the DOM, i.e., subjects with, respectively, zero and nonzero frequency parameter $\mu$.}
\end{center}
\end{table}

\begin{figure*}[!ht]
\includegraphics[width=8.7cm]{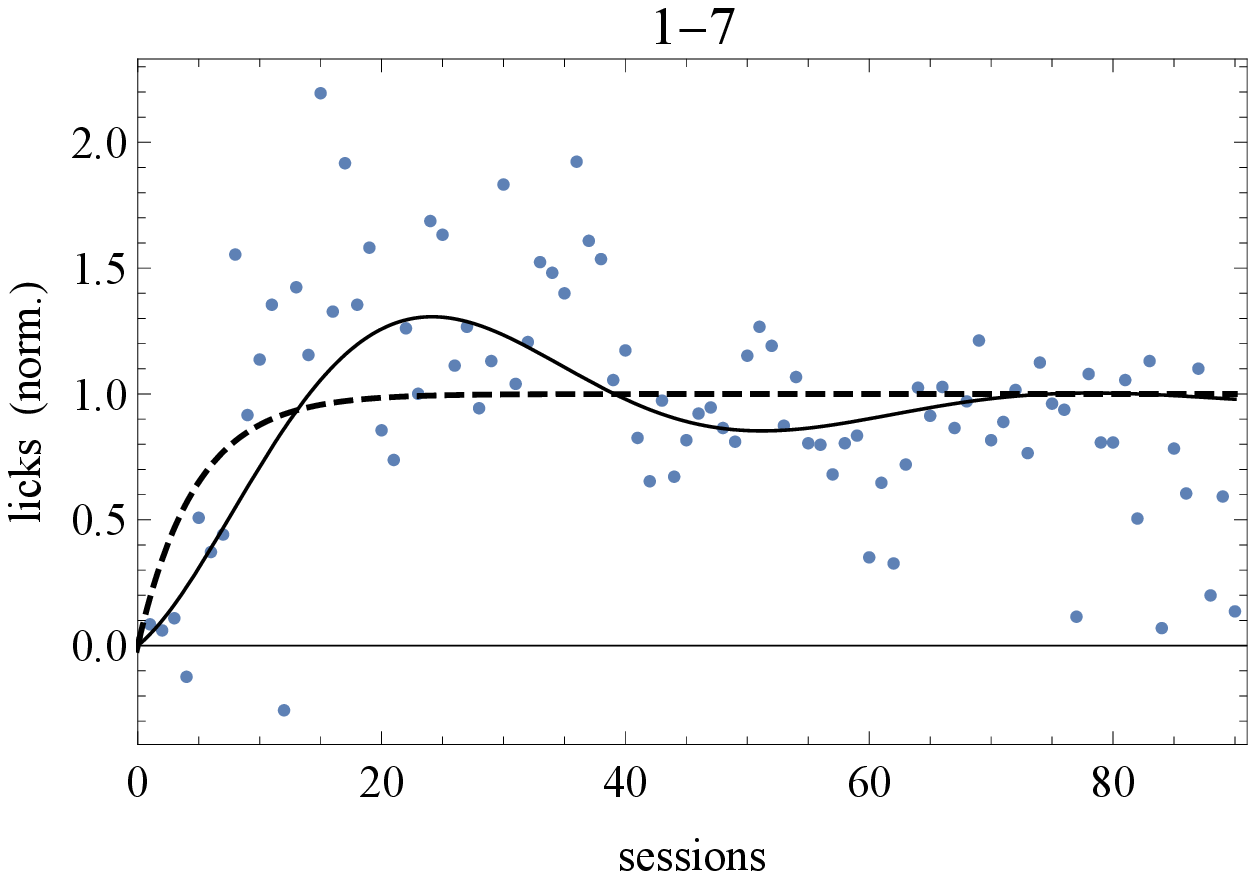}\includegraphics[width=8.7cm]{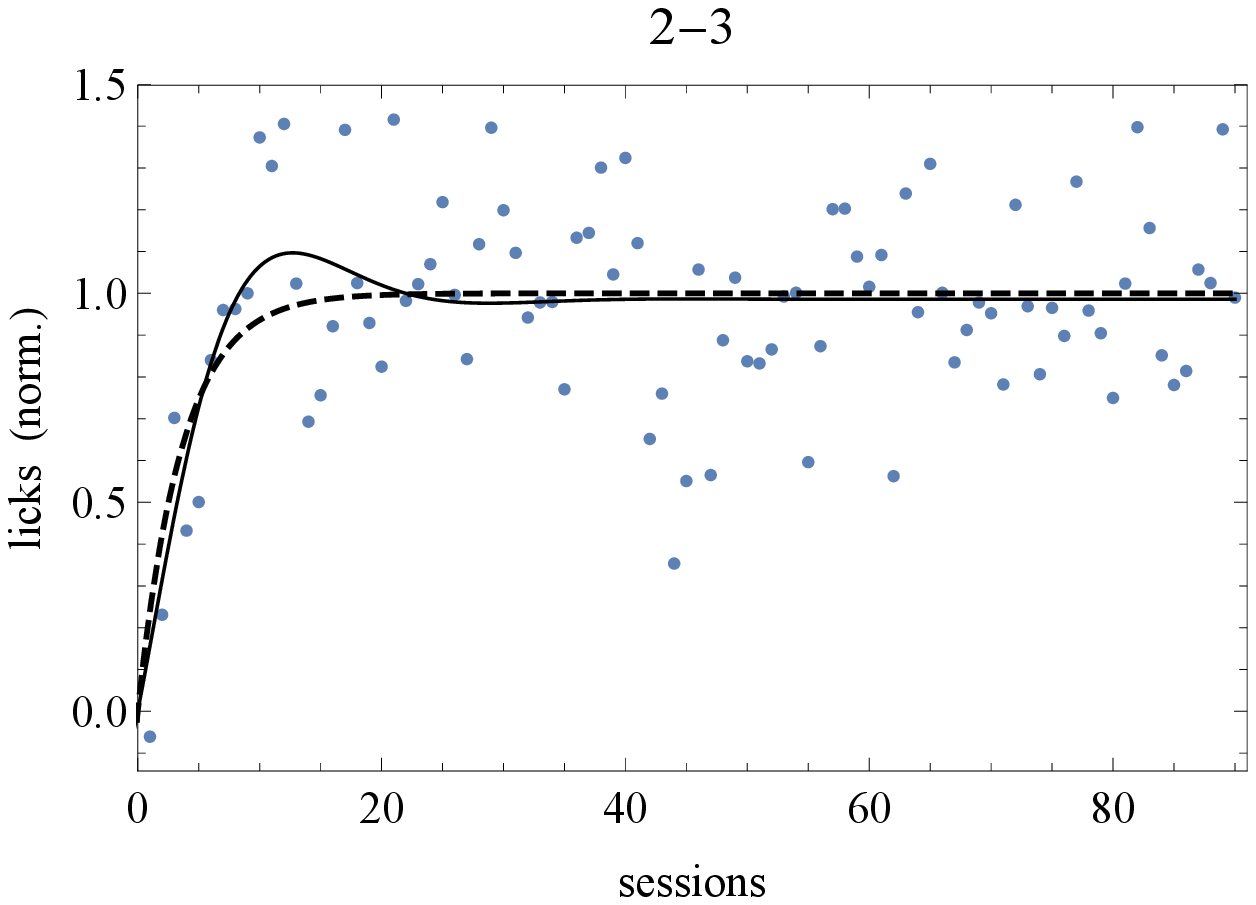}\\ \\
\includegraphics[width=8.7cm]{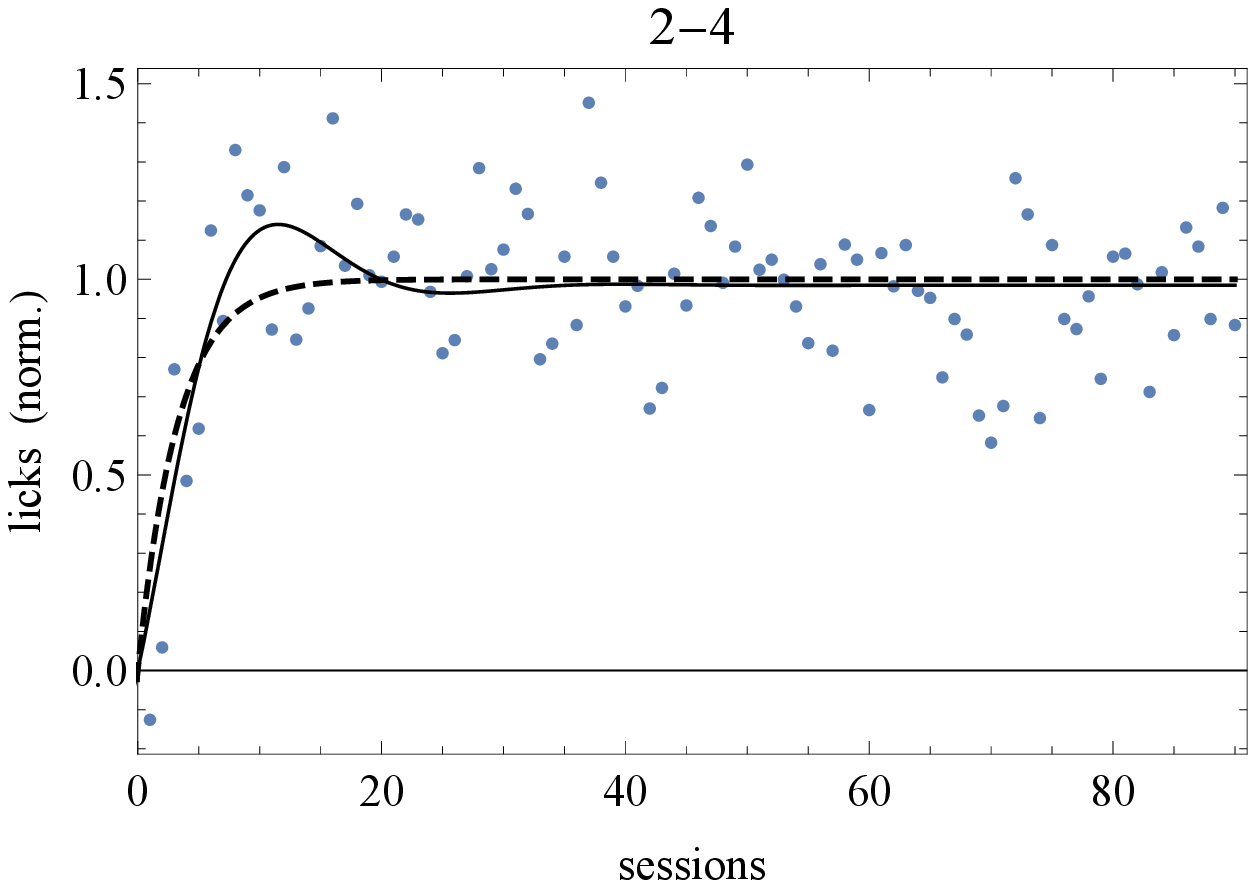}\includegraphics[width=8.7cm]{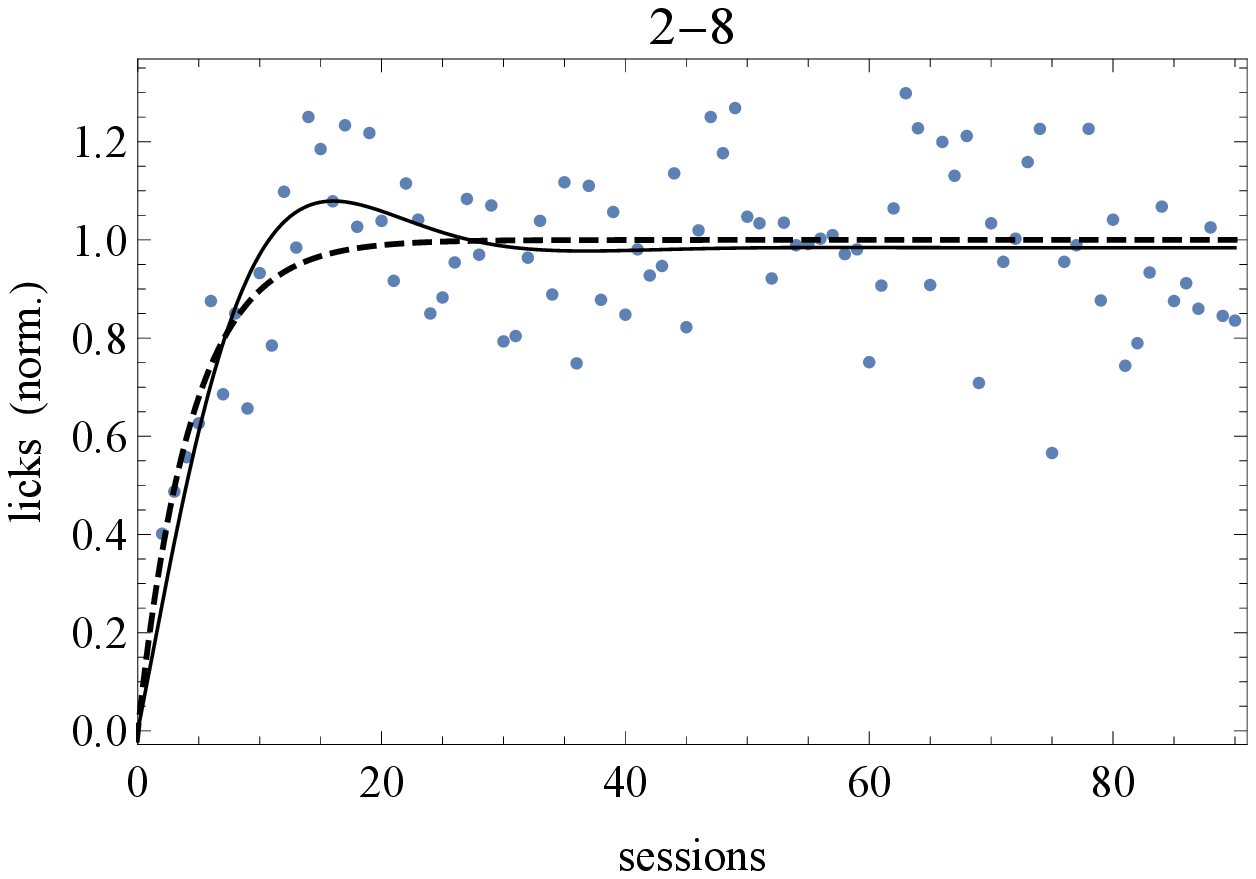}
\caption{\label{fig7} Best-fit of normalized data of our long experiment with the RW model \Eq{Hullearn} (dashed curve) and the DOM \Eq{oscso} with $A=0$ with the parameter values of {\bf Table \ref{tab4}}. Sessions are on the {\it horizontal axis} and normalized response is on the {\it vertical axis}.}
\end{figure*}

\bi
\item As one can see from the table, the RW best fit is favored for subjects 1-1, 1-2, 1-3, 1-4, 1-5, 1-6, 2-2, 2-6, and 2-7 (weak to positive evidence, $1\leq\De\leq 5$). However, here we are not comparing two independent models but a model and its extension by one ($\mu$) or two ($\mu,A$) extra parameters. Therefore, we have to interpret with care the meaning of the cases where RW is favored. We can divide them in three groups:
	\bi
	\item[--] Five cases (subjects 1-1, 1-2, 1-4, 1-5, 2-2, and 2-7) where also the DOM with mu close to zero fits the data. It means that for these subjects the RW model and the DOM are indistinguishable. For subjects 1-1, 1-2, 1-5, 2-2, and 2-7, the estimated $\mu$ in the oscillatory best fit is compatible with zero (1-1: $\mu=-0.13\pm0.17$; 1-2: $\mu=10^{-11}\pm 10^7$, 1-5: $\mu=6\pm 10^5$; 2-2: $\mu=10^{-6}\pm 10^4$; 2-7: $\mu=10^{-4}\pm 6$). Therefore, there is no statistically significant smooth oscillation of the type \Eq{oscso} in their response. For subject 1-4 (weak evidence in favor of RW according to the AIC, $\De=1$), $\mu$ is nonzero at the 68\% confidence level but zero at the 99\% confidence level ($\mu=0.45\pm 0.21$). 
	\item[--] Two cases (subjects 1-3 and 1-6) where the DOM fails to fit data unless one imposes $\mu=0$ by hand. Leaving this parameter free, here $\mu$ is significantly nonzero at the 99\% confidence level but is very high (1-3: $\mu=12.61\pm 0.04$; 1-6: $\mu=6.23\pm 0.02$), meaning that data are fitted with densely packed oscillations. We regard this as an artifact and thus discard these oscillatory fits as unviable. 
	\item[--] One case (subject 2-6) where the RW model is, so to speak, superior by brute force. Here $\mu$ is significantly nonzero at the 99\% confidence level and is not high ($\mu=0.04\pm 0.01$), but the BIC and AIC are larger than for the RW model. 
	\ei
From this discussion, we conclude that, of these nine cases, only one (subjects 1-3, 1-6 and 2-6) directly discards the DOM with $\mu\neq 0$, although with only weak to positive evidence. While in all the other cases Hull's primacy does not exclude the fact that, for these subjects, the DOM with $\mu$ close to zero is also allowed.
\item The data of subjects 1-7, 2-3, 2-4, and 2-8 favor the $A=0$ DOM with, respectively, very strong, moderate, strong, and moderate evidence. Restoring $A$ as a free parameters, we get an even better fit for subject 1-7. The $p$-value for $\mu$ in all these cases is very small, which means that the probability to have $\mu\neq 0$ by statistical chance is negligible.
\item The case of subjects 2-1 and 2-5 is less clear-cut and we have to look at the decimals: the $A=0$ DOM (2-1: $\mu=0.17\pm 0.03$; 2-5: $0.56\pm 0.04$) is slightly more favored than the RW model in the AIC for subject 2-5, and vice versa, so that we attribute one each. Since the difference is about 1\% of the value of the criteria and evidence in favor or against is weak ($\De=1$), this attribution is only for the sake of the final counting.
\ei
The best-fit parameter values of subjects 1-7, 2-3, 2-4, and 2-8 for the two models are given in {\bf Table \ref{tab4}} and the best-fit curves are shown in {\bf Figure \ref{fig7}}. Since the favored best fits with $A\neq 0$ are not particularly strong (the parameter $A$ is always zero at the 95\% confidence level), we can conclude that the model with $A=0$ is sufficient to fit successfully the data of these subjects deviating from the RW behavioral trend.
%4
\begin{table}[!ht]
\begin{center}
\begin{tabular}{lcccc}\hline
{\bf Subject}              & {\bf Parameters} & {\bf RW}          & {\bf Oscillations} & $\bm{p}$\\\hline
\multirow{5}{*}{1-7} & $\la$      & $1$             & $\mathbf{0.78\pm0.13}$ &\\
										 & $\a\b$     & $0.21\pm0.09$ & $\mathbf{0.04\pm0.01}$ &\\
										 & $\mu$      & $0$             & $\mathbf{0.05\pm0.02}$ & $<0.05$\\
										 & $A$        & $0$             & $\mathbf{-2.46\pm1.94}$&\\
										 & $\s$       & $0.43$          & $\mathbf{0.36}$ &\\\hline
%\multirow{5}{*}{2-1} & $\la$      & 1             & $0.97\pm 0.04$       & $0.98\pm0.04$\\
%										 & $\a\b$     & $0.22\pm0.07$ & $0.10\pm 0.03$       & $0.10\pm0.04$\\
%										 & $\mu$      & 0             & $0.17\pm 0.03$       & $0.19\pm0.05$\\
%  									 & $A$        & 0             & 0                    & $0.16\pm0.50$\\
%										 & $\s$       & 0.32          & 0.32								 & 0.32\\\hline	
\multirow{5}{*}{2-3} & $\la$      & $1$             & $\mathbf{0.99\pm 0.02}$ &\\
										 & $\a\b$     & $0.27\pm0.07$ & $\mathbf{0.15\pm 0.05}$ &\\
										 & $\mu$      & $0$             & $\mathbf{0.19\pm 0.04}$ & $<0.001$\\
										 & $A$        & $0$             & $\mathbf{0}$            &  \\
										 & $\s$       & $0.22$          & $\mathbf{0.22}$			& \\\hline
\multirow{5}{*}{2-4} & $\la$      & $1$             & $\mathbf{0.98\pm 0.02}$ &\\
										 & $\a\b$     & $0.30\pm0.07$ & $\mathbf{0.14\pm 0.03}$ &\\
										 & $\mu$      & $0$             & $\mathbf{0.22\pm 0.03}$ & $<0.001$\\
										 & $A$        & $0$             & $\mathbf{0}$        & \\
										 & $\s$       & $0.20$          & $\mathbf{0.18}$			& \\\hline
\multirow{5}{*}{2-8} & $\la$      & $1$             & $\mathbf{0.98\pm 0.02}$ &\\
										 & $\a\b$     & $0.23\pm0.03$ & $\mathbf{0.13\pm 0.02}$ &\\
										 & $\mu$      & $0$             & $\mathbf{0.15\pm 0.02}$ & $<0.001$\\
										 & $A$        & $0$             & $\mathbf{0}$ & \\
										 & $\s$       & $0.15$          & $\mathbf{0.15}$	&\\\hline
\end{tabular}
\caption{\label{tab4} Best-fit values of the parameters of the RW model \Eq{Hullearn} and the DOM \Eq{oscso} for those subjects whose data of our long experiment (normalized with respect to the RW asymptote) favor the DOM for both the BIC and the AIC (hence subject 2-5 is not shown). $\s$ is the estimated standard error. The $p$-value for $\mu$, the key parameter to distinguish the RW model from the DOM, is shown. The most favored model is in boldface.}
\end{center}
\end{table}
%7

\medskip

\subsubsection{Harris et al.\ (2015) Experiment 2}

\begin{figure}[!ht]
\includegraphics[width=8.7cm]{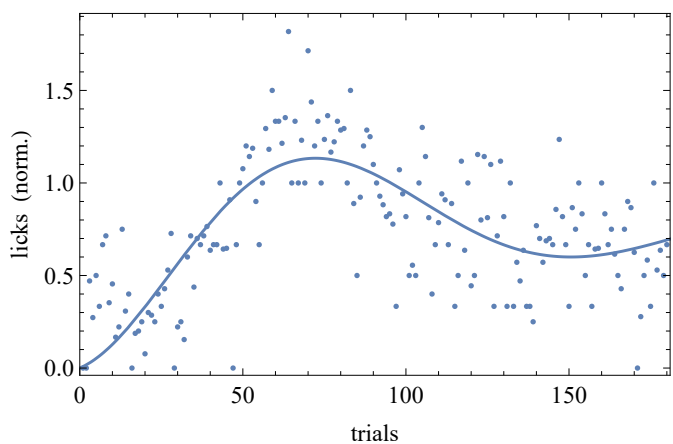}
\caption{\label{fig8} Trial-by-trial data of subject CR10-1 fitted with the DOM \Eq{oscso}. Trials are on the {\it horizontal axis} and response is on the {\it vertical axis}.}
\end{figure}

The BIC and AIC of the RW model and the DOM are shown in {\bf Tables \ref{tab5}}, {\bf \ref{tab6}} together with the difference $\De_{\rm HO}={\rm IC}_{\rm RW}-{\rm IC}_{\rm oscil}$. Favored models are in boldface; when two models are favored in different IC, the one with strongest evidence ``wins.'' Oscillatory fits yielding bigger IC and those that fail to produce a nontrivial model (i.e., if $\mu$ and/or $A$ vanish) are discarded. Trivial RW fits with vanishing $\a\b$ are reported except when also the other fits fail, in which case all cells are left blank. An example of fit is shown in {\bf Figure \ref{fig8}} and a summary of RW and oscillatory favored models is given in {\bf Table \ref{tab7}}.
%5
\begin{table}
\begin{center}
\begin{tabular}{lcccc}\hline
{\bf CR10} &	{\bf RW}    & {\bf Oscillations} & $\bm{\De_{\rm HO}}$ & $\bm{p\,(\mu)}$ \\\hline
%\multicolumn{4}{|c|}{CR10} \\\hline
1	&	$130$, $120$ & \textbf{55}, \textbf{43}  & $74$, $78$ & $<0.001$ \\
2	&\textbf{177}, \textbf{168}	&$182$, $170$ &	$-5$, $-2$ & \\
3	&\textbf{141}, \textbf{131} &	$146$, $133$ & $-5$, $-2$ & \\
4	&\textbf{119}, \textbf{109} &	$124$, $111$ &	$-5$, $-2$ & \\
5 &$99$, $90$ & \textbf{80}, \textbf{64}*	&	$19$, $26$ & \\
6	&$32$, $23$ &	\textbf{21}, \textbf{9}	& $11$, $14$ & $<0.001$\\
7	&$-58$, $-68$ &	{\bf $-$76}, {\bf $-$88} & $17$, $21$ & $<0.001$\\
8	&\textbf{141}, $132$ &	$142$, \textbf{130}	&	$-1$, $2$& $<0.001$ \\
9	&\textbf{140}, \textbf{130} &	$143$, $130$ & $-3$, $0$& \\
10&	\textbf{8}, {\bf $-$1}	& $13$, $1$	& $-5$, $-2$& \\
11&	161, 152 & \textbf{153}, \textbf{137}& 8, 15& $<0.001$\\
12&	\textbf{102}, \textbf{92} &	$107$, $94$	&	$-5$, $-2$& \\
13&	\textbf{82}, \textbf{72} &	$87$, $74$ &	$-5$, $-2$& \\
14&	$50$, $40$ & \textbf{49}, \textbf{36}	& $1$, $4$& $<0.001$\\
15&	$202$, $193$ & \textbf{193}, \textbf{177}	&	$9$, $16$& $<0.001$\\
16&	$144$, $134$ &	\textbf{143}, \textbf{130}& $1$, $5$& $<0.001$\\\hline
{\bf CR30} &	{\bf RW}    & {\bf Oscillations} & $\bm{\De_{\rm HO}}$ & $\bm{p\,(\mu)}$ \\\hline
%\multicolumn{4}{|c|}{CR30} \\\hline
1	&$149$, $139$&	\textbf{141}, \textbf{128}	&	$8$, $11$&$<0.001$\\
2	&\textbf{207}, \textbf{197}&	$211$, $198$	&	$-4$, $-1$&\\
3	&\textbf{23}, \textbf{13}&	$28$, $15$	&	$-5$, $-2$&\\
4	&$119$, $109$&	\textbf{117}, \textbf{104} &	$2$, $5$& $<0.001$\\	
5	&$115$, $106$&\textbf{42}, \textbf{26}*	&	$74$, $80$& \\
6	&			&	&& \\
7	&$-101$, $-111$&	{\bf $-$120}, {\bf $-$133}	&$19$, $23$& $<0.001$\\
8	&\textbf{163}, \textbf{154}&	$168$, $156$	&	$-5$, $-2$& \\
9	&\textbf{57}, \textbf{48}&	$63$, $50$	&	$-5$, $-2$& \\
10&	{\bf $-$83}, {\bf $-$92} &	$-78$, $-91$	&	$-4$, $-1$& \\
11&	103, 94&	\textbf{87}, \textbf{74}	&	$17$, $20$& $<0.001$\\
12&	92, 82&	\textbf{71}, \textbf{58}	&	$21$, $24$& $<0.001$\\
13&	{\bf $-$36}, {\bf $-$46} &	$-31$, $-44$	& $-5$, $-2$& \\
14&	$-87$, $-96$&	{\bf $-$94}, {\bf $-$107}	&	$8$, $11$& $<0.001$\\
15&	\textbf{96}, $86$& $102$, \textbf{86}	&	$-6$, $0$& \\
16&	\textbf{150}, \textbf{141}&	$156$, $143$	&	$-5$, $-2$& \\\hline
\end{tabular}
\caption{\label{tab5} BIC and AIC (in the format ``BIC, AIC'' and approximated to zero decimals) of the best fits of the trial-by-trial data of CR10 and CR30 groups in Experiment 2 of \citet{HPG} with the RW model \Eq{Hullearn} and the DOM \Eq{oscso}. The $p$-value for $\mu$, the key parameter to distinguish the RW model from the DOM, is shown when oscillations are favored. Favored models are in boldface. Although we fitted the DOM in two ways, one with $A=0$ from the start and one with $A$ free, we report only the best between these two versions. Asterisks mark possible false positives, for which $p\approx 1$ for $\mu$.}
\end{center}
\end{table}
%6
\begin{table}
\begin{center}
\begin{tabular}{lcccc}\hline
{\bf PR10} &	{\bf RW}    & {\bf Oscillations} & $\bm{\De_{\rm HO}}$ & $\bm{p\,(\mu)}$ \\\hline
%\multicolumn{4}{|c|}{PR10} & \\\hline
1&	$533$, $520$&	\textbf{450}, \textbf{433}&	$83$, $87$& $<0.001$ \\
2&	\textbf{458}, $445$&	$460$, \textbf{443}	&	$-2$, $2$& \\
3&	\textbf{513}, \textbf{500}&	$519$, $502$	&	$-6$, $-2$& \\
4&	$514$, $501$&	\textbf{482}, \textbf{465}	&	$31$, $36$	&  $<0.001$\\
5&	$416$, $403$&	\textbf{184}, \textbf{163}*	&	$231$, $240$& \\
6&					&&& \\
7&	\textbf{160}, \textbf{147}&	$166$, $149$	&	$-6$, $-2$& \\
8&	\textbf{483}, \textbf{470}&	$488$, $471$	&	$-5$, $-1$& \\
9&	\textbf{215}, \textbf{203}&	$222$, $205$	&	$-6$, $-2$& \\
10&	{\bf $-$75}, {\bf $-$88} &	$-71$, $-88$	& $-5$, $0$& \\
11&				&&	& \\
12&	$391$, $378$&	\textbf{368}, \textbf{351}	&	$23$, $28$&  $<0.001$\\
13&	\textbf{344}, \textbf{331}&	$350$, $333$	&	$-6$, $-2$& \\
14&	$25$, $12$&	\textbf{19}, \textbf{2}	&	$6$, $10$& $<0.001$\\
15&	\textbf{439}, \textbf{426}&	$445$, $428$	&	$-6$, $-2$& \\
16&	\textbf{581}, \textbf{568}&	$587$, $570$	&	$-6$, $-2$& \\\hline
{\bf PR30} &	{\bf RW}    & {\bf Oscillations} & $\bm{\De_{\rm HO}}$ & $\bm{p\,(\mu)}$ \\\hline
%\multicolumn{4}{|c|}{PR30} \\\hline
1&	\textbf{301}, $288$& $309$, \textbf{287}	&	$-8$, $1$& \\
2&	$555$, $542$&	\textbf{533}, \textbf{516}	&	$22$, $26$	& $<0.001$\\
3&	\textbf{334}, \textbf{322}&	$341$, $324$	&	$-6$, $-2$& \\
4&	$316$, $303$&	\textbf{310}, \textbf{293}	&	$6$, $10$	& $<0.001$\\
5&	\textbf{129}, \textbf{116}&	$136$, $118$	&	$-6$, $-2$& \\
6&	$-343$, $-356$&	{\bf $-$350}, {\bf $-$367} &		$7$, $11$	& $<0.001$\\
7&	$-199$, $-212$&	{\bf $-$217}, {\bf $-$234} &	$18$, $22$& $<0.001$\\
8&	$244$, $231$&	{\bf 230}, {\bf 208}	&	$14$, $22$& $<0.001$\\
9&	\textbf{109}, \textbf{96}&	$115$, $98$&	$-6$, $-2$	& \\
10&	{\bf $-$190}, $-202$&	$-184$, {\bf $-$206} &	$-5$, $3$& \\
11&	$211$, $198$&	\textbf{62}, \textbf{40}	&	$150$, $158$& $<0.001$\\
12&	$208$, $196$&	\textbf{202}, \textbf{185}	&	$7$, $11$& $<0.001$\\
13&	$28$, $15$&	\textbf{26}, \textbf{9}	&	$2$, $6$& $<0.001$\\
14&	$-174$, $-187$&	{\bf -201}, {\bf $-$223} &	$27$, $36$& $<0.001$\\
15&	$311$, $298$&	{\bf 296}, {\bf 275}	&	$15$, $24$	& $<0.001$\\
16&	\textbf{724}, $711$&	$724$, \textbf{707}	&	$0$, $4$	& $<0.001$\\\hline
\end{tabular}
\caption{\label{tab6} BIC and AIC (in the format ``BIC, AIC'' and approximated to zero decimals) of the best fits of the trial-by-trial data of PR10 and PR30 groups in Experiment 2 of \citet{HPG} with the RW model \Eq{Hullearn} and the DOM \Eq{oscso}. The $p$-value for $\mu$, the key parameter to distinguish the RW model from the DOM, is shown when oscillations are favored. Favored models are in boldface. Although we fitted the DOM in two ways, one with $A=0$ from the start and one with $A$ free, we report only the best between these two versions. Asterisks mark possible false positives, for which $p\approx 1$ for $\mu$.}
\end{center}
\end{table}
%7
\begin{table}
\begin{center}
\begin{tabular}{lccccc}\hline
     & {\bf CR10 (\%)} & {\bf CR30 (\%)} & {\bf PR10 (\%)} & {\bf PR30 (\%)} & {\bf Total (\%)} \\\hline
RW & 44 & 50 & 56 & 31 & 45\\
Oscillatory & 56 & 44 & 31 & 69 & 50\\
No fit &  & 6 & 13 &  & 5\\\hline
\end{tabular}
\caption{\label{tab7} Percentage of subjects of Experiment 2 of \citet{HPG} following the RW model or the DOM, i.e., subjects with, respectively, zero and nonzero frequency parameter $\mu$. The learning curve of subjects CR30-6, PR10-6, and PR10-11 did not follow any of the models.}
\end{center}
\end{table}
%8

A group comparison using {\bf Tables \ref{tab5}--\ref{tab7}} shows that:
\bi
\item When the RW model is favored, it is so with weak to positive evidence in CR groups and positive to strong evidence in PR groups. This increase in evidence in favor (but not in the number of favored cases) is probably due to the larger number of data points (three times as many in PR groups with respect to CR groups).
\item On the other hand, the great majority of favored DOMs have very strong evidence in favor. Thus, when subjects display oscillations in their learning curve the effect is usually strong, not just tiny perturbations of the monotonic RW curve.
\item Within the same type of reinforcement (continuous or partial), when extending the trial duration this percentage switches in favor of the other model, but in the opposite way of the groups with the other type of reinforcement (CR10 and PR30: majority is oscillatory; CR30 and PR10: majority is RW). This puzzling pattern may have an explanation, which we support by the data in section 2.4 of the \textbf{Supplementary Material}. The bottom line is that longer trials stabilize the behavior, but a partial reinforcement schedule destabilizes it, where increasing stability means fewer erratic patterns and/or more RW patterns, without being accompanied by smaller oscillations in oscillatory patterns. We will confirm this finding in \citet{CHP}.
\ei

%%%%%%%%%%%%%%%%%%%%%%%%%%%%%%%%%%%%%%%%%%%%%%%%%%%%%%%%%%%%%%%%%%%%%%%%%%%%%%

\medskip

\subsection{Discussion}

In our long experiment, the RW model is favored (but only moderately) with the respect to the $\mu\neq 0$ DOM for only one out of the 15 experimental subjects; the DOM with $A=0$ and $\mu\neq 0$ is strongly favored in two subjects and moderately favored in two more; the DOM with $A=0$ and $\mu=0$ coincides with the RW model and is a good fit of eight more subjects; and there is no preference for either the RW model or the DOM in two more subjects. The fact that the RW model is favored (and moderately so) only in 7\% of the experimental subjects with the respect to the $\mu\neq 0$ DOM justifies the present and future interest in the $A=0$, $\mu\neq 0$ DOM \Eq{oscso}.

Since RW is a subcase of the DOM, if RW is favored for a subject, then the DOM with frequency approximately zero may also be a good fit, while if one of the DOMs is favored, it means that the frequency cannot be set to zero and RW is not a good fit. Since the goal is to fit as many subjects as possible with the same model, then the conclusion from our experiment ({\bf Table \ref{tab3}}) is that, if we insist in considering the RW model as the correct description of the learning curves, then we can explain at most 67\% of the data ($6+4=10$ subjects out of 15), while if we postulate the least-action principle we obtain a model that includes RW as a special case and can fit successfully $100\%$ of the data. Also the results from Experiment 2 of \citet{HPG} ({\bf Table \ref{tab7}}) leads us to conclude that the DOM provides a more powerful explanation of the data than RW, since it can account for 95\% of the data against the 45\% by RW. Therefore, this extension of the RW model is both natural and viable.

%%%%%%%%%%%%%%%%%%%%%%%%%%%%%%%%%%%%%%%%%%%%%%%%%%%%%%%%%%%%%%%%%%%%%%%%%%%%%%
%%%%%%%%%%%%%%%%%%%%%%%%%%%%%%%%%%%%%%%%%%%%%%%%%%%%%%%%%%%%%%%%%%%%%%%%%%%%%%

\section{Stochastic model of individual behavior}\label{noises}%\section{COLORED STOCHASTIC MODEL OF INDIVIDUAL BEHAVIOR}

%%%%%%%%%%%%%%%%%%%%%%%%%%%%%%%%%%%%%%%%%%%%%%%%%%%%%%%%%%%%%%%%%%%%%%%%%%%%%%

\medskip

\subsection{Motivation and spectral analysis}

The DOM describes long-range modulations in the subjects' responses, where an oscillation period (a peak followed by a trough) spans tens of sessions. However, even the most cursory look at the data reveals that response variates somewhat wildly at a much shorter scale, from session to session or even from trial to trial. In this section, we will study these short-scale variations quantitatively using a spectral analysis and we will connect our results with an interesting open question in the cognitive literature: ¿Should these fluctuations be treated as experimental error or do they have an origin in the cognitive modules of the subject? While the answer seems to be No--Yes when subjects are human, little or no attention has been drawn to the case of animals. According to our results, the answer is Yes--No: response fluctuations occurring at short time scales are indistinguishable from experimental error.

The generalized presence of large short-scale fluctuations in the response translates into an unstable inter-trial and inter-session associative strength. These fluctuations are present in \citet{HPG} experiments as well as in ours, even in those very few subjects with relatively small inter-session variability. %, such as 2-7.
 Since there seems to be no qualitative change with the time scale (see {\bf Figure S18}), fluctuations might be assumed to plague the performance of the subjects at \emph{any} scale. This means that it is quite natural to regard these data as a time series described by a nowhere-differentiable pattern instead of a smooth learning curve. This is the motivation to replace \Eq{DeV} with an evolution equation for the associative strength encoding a stochastic component.\footnote{The mathematics of stochastic processes is not a new tool in learning theories. In fact, the earliest conditioning models had an inherent element of randomness in their predictions. The classic 1950s models were cast in the language of probability theory and one considered the probability $p$ of a given conditioned response (CR) as a function of the trial number $n$; see, e.g., the works by Estes \citep{Est50,EsBu1}, Bush and Mosteller \citeyearpar{BuMo1,BuMo2,BuMo3}, and the reviews by \citet{Mos58} and \citet{Bow94}. Later on, the strength of association $v$ was regarded as a better operational variable than the probability $p$ \citep{RW72} and the focus was shifted to ideally deterministic predictions. More recently, stochastic processes played a role in the context of neural networks and their application to robotics and artificial intelligence. In particular, a path integral can describe a learning process of a neural network as a finite-temperature stochastic process \citep{Bal00}. Path integrals have also been applied to control theory and reinforcement learning (\citealp{BOTS}; \citealp{FaBu}; \citealp{Kap07}; Pan et al., \citeyear{PTK1,PTK2}; \citealp{The11}; Theodorou et al., \citeyear{TBS1,TBS2,TSBS}; \citealp{VWK}). In all these cases, the problem is to minimize the cost of a learning process or of an action by an agent. Despite some remote similarities, these approaches differ from ours.}

A \emph{spectral analysis} is an analysis of the frequency modes constituting the noise in data. Technically, it amounts to take the Fourier transform of the data. Intuitively, it means to count how many times a response fluctuation of a given amplitude occurs for a subject during the experiment. If data predominantly oscillate on a long time scale, then the noise dominates the small frequencies of the spectrum. On the other hand, if oscillations mainly occur on a very short time scale, then the spectrum is more noisy at large frequencies. The noise in learning data corresponds to the second case.

In essence, the variability of the response of a subject is encoded in the \emph{frequencies} $\om$, distributed in the so-called \emph{power spectral density} $S(\om)$, obtained from the response. If the response is perfectly stable from trial to trial, then all frequencies are equally represented in $S(\om)$ and the power spectral density is constant. If the response varies gradually as in the ideal learning curve of the RW model, then $S(\om)$ is a smooth function. If, however, the response varies erratically from trial to trial, as in actual data, then $S(\om)$ becomes very ragged and this raggedness is what we call \emph{noise}. The basics of spectral analysis and theoretical shape of the power spectral density for the RW model without and with noise are presented in the \textbf{Supplementary Material}. Here we only note that noise alone is usually represented by the power spectral density
\be\label{noise}
S_a(\om)=\frac{1}{\om^a}\,,
\ee
where $a$ is a constant. When $a=0$, noise is said to be \emph{white}, while it is \emph{colored} if $a\neq 0$.

The nature of the noise source in data can be discriminated by the large-frequency region of the power spectral density. If all noise comes from statistical error, then this region is flat in the log-log plane. However, if this region is not flat and exhibits a positive or negative average slope, then it is possible that some other random noise-generating mechanism, of cognitive-behavioral origin, is in action.

%%%%%%%%%%%%%%%%%%%%%%%%%%%%%%%%%%%%%%%%%%%%%%%%%%%%%%%%%%%%%%%%%%%%%%%%%%%%%%

\medskip

\subsection{Theory}

The RW model with colored noise admits a straightforward mathematical treatment as a stochastic process, where $v_n$ is promoted to a random variable $V_n=v_n+\xi_n$, where $v_n$ and $\xi_n$ are, respectively, the deterministic and stochastic parts (see \textbf{Supplementary Material}).

To give $\xi_n$ a psychological interpretation, we can draw inspiration from the ``$1/f$ noise'' cognitive literature, where the variability in human response in memory tasks, reaction tasks, mental rotation, word naming, and so on, is characterized by a colored noise (Dixon et al., \citeyear{DHMS,DSBA}; \citealp{EHKK}; \citealp{FWR1}; Gilden, \citeyear{Gil01}; Gilden et al., \citeyear{GiTM}; Holden, \citeyear{Hol05,Hol13}; Holden et al., \citeyear{HVOT,HCAV}; Kello et al., \citeyear{KBHVO,KAHVO}; Ihlen and Vereijken, \citeyear{IV1,IV2}; \citealp{LFAA}; Stephen et al., \citeyear{SBMD,SDI}; Van Orden et al., \citeyear{VOHT1,VOHT2}; \citealp{TG}; Wagenmakers et al., \citeyear{WFR1,WFR2,WvF}; see \citealp{Kel10}, and \citealp{RiHo}, for reviews). There are three interpretations of this phenomenon. In the \emph{idiosyncratic} view, this noise is regarded as the intrinsic uncertainty, possibly due to an internal estimation error, in the formation of mental representations, such as the reproduction of spatial or temporal intervals in human memory \citep{GiTM}. Different cognitive systems combine to accidentally generate an overall noise term (Farrell et al., \citeyear{FWR1}; Wagenmakers et al., \citeyear{WFR1,WFR2,WvF}). In the \emph{nomothetic} view, the stochastic component is a more fundamental mechanism where cognition emerges as a self-organizing complex dynamical system (Dixon et al., \citeyear{DHMS,DSBA}; Gilden, \citeyear{Gil01}; Gilden et al., \citeyear{GiTM}; Holden, \citeyear{Hol13}; Ihlen and Vereijken, \citeyear{IV1}; Riley and Holden, \citeyear{RiHo}; Stephen et al., \citeyear{SBMD,SDI}; Van Orden et al., \citeyear{VOHT1}). In particular, the colored noise is the collective expression of the metastable coordination of different cognitive and motor systems in the performance of a task \citep{KBHVO}.  A third view intermediate between the idiosyncratic and the nomothetic was also considered \citep{LFAA}. 

Translating these considerations to the realm of non-human animal behavior, we can entertain the possibility \citep{Cal1} that, if nonwhite noise were detected in experiments of Pavlovian or operant conditioning, one would be observing a signal coming form the coordination of motor systems with the internal functioning of the subject's mind (either as an averaging of multiple cognitive subsystems or as an emergent collective manifestation of such coordination). We will come back to this cognitive perspective after analyzing the data.

%%%%%%%%%%%%%%%%%%%%%%%%%%%%%%%%%%%%%%%%%%%%%%%%%%%%%%%%%%%%%%%%%%%%%%%%%%%%%%

\medskip

\subsection{Data analysis: comparison with the RW model}

Session-by-session data are a coarse-grained version of the full data set of trial points. Inevitably, this coarse graining can hide or distort stochastic signals present at all time scales. For this reason, we consider trial-by-trial data. It is not difficult to check that a similar analysis done with session-by-session data yield the same results, but with a much greater error in the fits.

Note that the fits of the oscillatory and RW models were made with session data (90 points) for our long experiment, while those for the data of \citet{HPG} were made with trial data (180 or 540 points). Trial data of our experiment have larger dispersion than \citet{HPG} trial data, while their dispersion is comparable when the former data are binned into sessions.

\medskip

\subsubsection{Our experiment}

%Table \ref{tab5} reports the parameters of the RW model fitting trial-by-trial data, while 

The \textbf{Supplementary Material} includes the parameters of the RW model fitting trial-by-trial data and the power spectral density of the signal of the experimental subjects, calculated from each individual set of trial data points.

We look for a fit of the type of \Eq{noise} (times a constant which plays no role here) of the region of the spectral densities dominated by the stochastic noise, typically for $\om>0.1$. Fitting the power spectra from $\om=0.1$ to $\om=30$, we get that %the results of Tab.\ \ref{tab6}.
 the parameter $a$ is zero within the experimental uncertainty at the $1\s$-level for all subjects except 1-1, 1-3, and 2-3, where $a$ vanishes at the $2\s$-level. In all cases, the best-fit value of $a$ is very close to zero, at least in one part over one hundred. Therefore, all subjects of both groups display white noise.

Note that the spectral analysis of the noise signal occurs in a frequency region unaffected by whether the background model is RW or the oscillatory one, Eqs.\ \Eq{Hullearn} and \Eq{oscso}. %The power spectral density of the latter would only differ in the position of the right end of the plateau in {\bf Figure \ref{fig13}} and in an extra bump just at the onset of the slope. When fitting high-frequency data, these details are subdominant with respect to the main noise trend and, in fact, they do not appear in {\bf Figure \ref{fig16}.

\medskip

\subsubsection{Harris et al.\ (2015) Experiment 2}

Also for this experimental set the spectral noise function fit was done from a minimal to a maximal frequency, with a given sampling rate. The maximal frequency was chosen to be 60 Hz because the time resolution of recorded data in \citet{HPG} was of 16.6 ms. As a lower frequency, we took 1 Hz because at frequencies lower than that the spectrum is dominated by the learning curve (i.e., the spectrum becomes a deterministic smooth curve at about $\om<0.5-1$ Hz for all subjects). Including smaller frequencies would introduce artifacts. Also, trial lengths do not extend beyond a certain value (18 s for the PR CSs and 58 s for the CR CSs), which puts a limit on how low the frequency can be. A sampling rate of $\om=1/58=0.02$ Hz for the trials of up to 58s and $\om=1/18=0.05$ Hz for trials up to 18s was used. 

The values of $a$ of the noise best fits are all very small (see \textbf{Supplementary Material}) and the conclusion is that noise is white for all subjects at the $2-3\s$-level. All groups show the same noise in average.

%%%%%%%%%%%%%%%%%%%%%%%%%%%%%%%%%%%%%%%%%%%%%%%%%%%%%%%%%%%%%%%%%%%%%%%%%%%%%%

\medskip

\subsection{Discussion}

Having checked the presence of white noise in individual subjects, we turn to the psychological interpretation of these results. From a strictly behaviorist point of view, the question of whether this noise comes from a naive sum over different cognitive systems or arises as an emergent phenomenon is immaterial. These data cannot tell us anything about either alternative. However, we observe the same noise in all subjects and this noise is \emph{white}. The simplest explanation is that its origin is statistical and implies no fundamental property of the ``rat mind.'' If there were an underlying motor-cognitive mechanism (naive interference or emergent phenomenon) depending on the individual, on the task, and on the relevance of the stimuli for the subject and its learning history, one would see these differences in the noise trend. Although we cannot exclude this possibility, our data do not yield support to it. In particular, they exclude the cognitive function of \emph{attention} as the main responsible for colored noise. In fact, the main processes involved in this Pavlovian experiment are motor and attentional, and the only source of a colored signal could come from attention alone (or its interference with motor processes; idiosyncratic view) or as an emergent phenomenon from the combination of attention and motor processes (nomothetic view). The conclusion is that higher-order cognitive functions, absent in this experiment but present in those of human response, may be the main source of colored noise.

Talking about response variability in the learning curve, \cite{Gal04} conjectured that data of conditioning experiments might bear a trace of a colored noise. We do not confirm this conjecture here and, on the other hand, we also insist on the existence, at a significant confidence level, of a learning asymptote, which was questioned by the same authors.

We also stress that the cognitive-noise hypothesis is descriptive but not predictive. To the best of our knowledge, no explicit model of the ``internal working of the mind'' predicting the observed diversity of noise spectra has been proposed. In the \emph{Supplementary material}, we  advance a quantitative theory based on quantum mechanics giving this type of prediction.

%%%%%%%%%%%%%%%%%%%%%%%%%%%%%%%%%%%%%%%%%%%%%%%%%%%%%%%%%%%%%%%%%%%%%%%%%%%%%%
%%%%%%%%%%%%%%%%%%%%%%%%%%%%%%%%%%%%%%%%%%%%%%%%%%%%%%%%%%%%%%%%%%%%%%%%%%%%%%

\section{Conclusions}\label{geco}%\section{GENERAL CONCLUSIONS}

By assuming minimization of the action as a guiding principle, we have constructed a new framework of models of Pavlovian conditioning. The simplest of these models is a natural extension of Rescorla--Wagner model with one cue (RW model in short). Looking at data of two experiments, we saw that the oscillatory classical model \Eq{oscso} with $\mu\neq 0$ is favored over RW in 42\% of the experimental cases. The estimated dispersion of the session-by-session individual data with respect to the theoretical curve is about the same for the RW and DOMs in all the cases where the latter is clearly favored. In all the other cases, the parameter $\mu$ is close to zero and the RW model and its extension coincide. Almost all (98\%) data can be fitted successfully by the DOM, which is an extension of the RW model, while RW alone can account only for 56\%. One may wonder whether this encouraging outcome was due to the presence of oscillations \emph{per se} rather than for an intrinsic virtue of the new model. This is not the case, as we checked in a separate publication \citep{CHP}.

At a biological level, the explanation of why $\mu$ takes different values in different subjects (for some, as we have seen, $\mu$ is close to zero) might reside in individual differences in brain configuration. Different dynamical adaptations of synaptic connections during learning would give rise to a different damping rate of the oscillations. Regardless of whether this microscopic interpretation has a basis in reality or not, the least-action principle states that learning is a naturally effective process that minimizes the internal changes in the subject. These changes are observable and can be quantified through the behavioral laws stated here.

In parallel, using individual trial-by-trial data we have checked for the presence of colored noise in the frequency spectrum of the subjects' response. We found no evidence of color and all subjects showed a flat (white) noise spectrum. It is important to stress the difference between the molecular and molar levels of analysis we considered in this paper: the spectral analysis probes response fluctuations at a much shorter time scale than learning models such as the RW model and the DOM. While the first analysis relies on a purely stochastic pattern, the latter works on a continuum or quasi-continuum. This means, in practice, that short-scale deviations from the asymptote are pure noise, while long-range deviations have, according to evidence, an associative oscillatory nature. The DOM is not just a spurious fit to the noise.

The origin of this white noise can be simply statistical, but we also considered two quantitative models accounting for it. One is a descriptive model implementing stochastic fluctuations in the subject's individual response; speculations about the origin of this noise may find inspiration in the ``$1/f^\a$'' cognitive literature, albeit in that case there is established evidence of a colored spectrum. The other model (see \textbf{Supplementary Material}) is not only descriptive, but also predictive, and interprets response variability as a manifestation of a process obeying the laws of quantum mechanics. We tested the predictions of the theory with data and we found agreement, although the error bars on the estimates of the ``Planck'' constant $\bar h$ are too conspicuous to conclude that the model is correct.

Although data show that individual responses are far from following the textbook smooth learning curve, they do not show an ``abrupt acquisition'' phenomenon either, as sometimes claimed in the literature (see \citealp{GDB}, and references therein). Some subjects do show something that could be described as an abrupt acquisition, but response variability is too large throughout the experiment to make this conclusion meaningful when, to put it simply, a sharp initial rise in response is yet another random fluctuation around the ideal average curve (see {\bf Figures S4} and {\bf S6}). Precisely for the same reason, although it is true that traditional associative models do not predict abrupt chances in behavior, our findings do not support representational or model-based models either \citep{GDB}. Response variability is so fine grained that its random, nowhere-differentiable nature is, in our opinion, unquestionable. We have offered two interpretations about these behavioral fluctuations, stochastic (from cognitive noise component in the underlying model) or quantum (from a model affected by quantum uncertainties). Both cases go beyond standard associationism, but are based on it nevertheless: associative learning still is an adequate description of averaged data.

Regarding replicability and applicability, since all the models we have presented in the main text of this paper are foundational extensions of the \emph{simplest} conditioning process involving simple associations between stimuli and responses, it may not be necessary to conduct \emph{ad hoc} experiments to test their validity. The long-range oscillations of the dynamical model, the spectral properties of response variations, and all the main features the quantum model can be checked in any experiment, past or future, that had a sufficient number of sessions (in the case of the dynamical model) or trials (in the case of the stochastic noise and of the quantum model) and whose design induced a simple conditioning process. In general, the condition of having many sessions is much more restrictive than that of having many trials, but it is not necessary to take or make experiments as lengthy as ours. Eventually, what one is looking for is cumulative evidence, and that is achievable with enough experiments of moderate length. We made the point in this paper by reanalyzing the data of an experiment first presented by \citet{HPG}. Moreover, our models may obtain confirmation or be ruled out by extant or future data not only about Pavlovian conditioning, but also from the operant conditioning literature. In the second case, one invokes the possibility that associative models can also describe operant behavior \citep{KN18}. In the first case, the generality of the construction of learning models as dynamical models and the psychological interpretation put forward in section \ref{secthe} make us believe that they can both be extended to other Pavlovian conditioning paradigms, such as eyeblink conditioning, where the average learning pattern may differ from the RW curve. The main point is not only that recasting a learning model as a dynamical model should always be possible, but that in doing so one can discover interesting modifications dictated by the naturalness argument (absence of fine tuning), such as the introduction of a new parameter $\mu$ in the DOM case. An oscillatory behavior may or may not arise from this construction, but this is something to be checked \emph{a posteriori}, not imposed by hand.

% We checked this possibility explicitly by inventing an ad-hoc \emph{non-associative oscillatory model} (NAOM) where sine and cosine oscillations are superposed to the RW learning curve \citep{CHP}. This model is non-associative because oscillations stem from a degree of freedom completely independent of the association strength and its purpose is to fit data better than the more constrained DOM discussed here. Using the same data considered here, it turns out that the NAOM can fit data better than the RW model, but no better than the DOM. In other words, the NAOM depletes the pool of subjects we here attributed to RW, while leaving the pool attributed to DOM virtually untouched. Overall, we found up to very strong evidence that the dynamical oscillatory model can fit more individual data than the RW model and the NAOM. The NAOM and the related model-selection analysis will be presented in detail in a separate publication \citep{CHP}.

%%%%%%%%%%%%%%%%%%%%%%%%%%%%%%%%%%%%%%%%%%%%%%%%%%%%%%%%%%%%%%%%%%%%%%%%%%%%%%
%%%%%%%%%%%%%%%%%%%%%%%%%%%%%%%%%%%%%%%%%%%%%%%%%%%%%%%%%%%%%%%%%%%%%%%%%%%%%%

\medskip

\paragraph*{Open Practices Statement} \ \noindent The data for all experiments are available upon request to the authors, and none of the experiments was preregistered.

\medskip

\paragraph*{Data Availability Statement} \ \noindent The datasets generated for this study are available on request to
the corresponding author.

\medskip

\paragraph*{Ethics Statement} \ \noindent The animal study was reviewed and approved by UNED bioethics.

\medskip

\paragraph*{Author Contributions} \ \noindent GC wrote the main body of the paper. EC-G and RP contributed
to editing the text. All the authors participated in the execution of one of the experiments.

\medskip

\paragraph*{Acknowledgments} \ \noindent GC thanks Pedro Vidal for invaluable help in writing the PASCAL programs of the pilot experiments that eventually led to the program used in this paper and Esmeralda Fuentes, Gabriela E.\ López-Tolsa, Ana de Paz and Valeria E.\ Gutiérrez-Ferre for rat-related advice. All authors thank Justin Harris for many fruitful discussions and for giving us access to the data of \citet{HPG}, Ralph Miller for his comments and suggestions about the experimental design, Antonio Rey for his outstanding handling of the lab schedule and of technical issues, and Vanessa Roldán for her involvement in the experiment and daily lab routine through part of this project. The experiment was run at the Animal Behavior Lab, Departamento de Psicología Básica I, Facultad de Psicología, Universidad Nacional de Educación a Distancia, Madrid, Spain, and was supported by grant PSI2016-80082-P from Ministerio de Economía y Competitividad, Secretaría de Estado de Investigación, Desarrollo e Innovación, Spanish Government (RP).

\medskip

\paragraph*{Supplementary material} \ Additional theoretical material (discussions on the main model, new models, and so on), figures, and tables are available in the \href{}{\textbf{Supplementary Material}}.

\medskip

\paragraph*{Conflict of Interest Statement} \ \noindent The authors declare that the research was conducted in the absence of any commercial or financial relationships that could be construed as a potential conflict of interest.

%%%%%%%%%%%%%%%%%%%%%%%%%%%%%%%%%%%%%%%%%%%%%%%%%%%%%%%%%%%%%%%%%%%%%%%%%%%%%%
%%%%%%%%%%%%%%%%%%%%%%%%%%%%%%%%%%%%%%%%%%%%%%%%%%%%%%%%%%%%%%%%%%%%%%%%%%%%%%

\bibliographystyle{apacite}

\end{document}